\documentclass[
reprint,
superscriptaddress,
showpacs,
nofootinbib,
amsmath,amssymb,
aps,
prl,
floatfix,longbibliography
]{revtex4-2}

\makeatletter
\let\auto@bib@innerbib\@empty
\makeatother

\usepackage{graphicx}
\usepackage{dcolumn}
\usepackage{booktabs}
\usepackage{mathtools}
\usepackage{bbm}
\usepackage{bm}
\usepackage{mathrsfs}
\usepackage{xcolor}
\usepackage{siunitx}
\usepackage{appendix}
\usepackage{makecell}
\usepackage{placeins}
\usepackage{array}

\usepackage[normalem]{ulem}

\usepackage[
hypertexnames=true,
breaklinks=true,
bookmarksnumbered=true,
bookmarksopen=true,
colorlinks=true,
linktocpage=true,
citecolor=blue,
urlcolor=magenta,
linkcolor=magenta
]{hyperref}

\newcommand{\bra}[1]{\langle #1|}
\newcommand{\ket}[1]{|#1\rangle}

\begin{document}

\title{Emergent Kinetic Constraints and Subspace Fragmentation in Rydberg Arrays}

\author{Wen-Jie Geng}
\thanks{These authors contributed equally to this work.}
\affiliation{Laboratory of Quantum Information, University of Science and Technology of China, Hefei 230026, China}

\author{Zhenming Zhang}
\thanks{These authors contributed equally to this work.}
\affiliation{Laboratory of Quantum Information, University of Science and Technology of China, Hefei 230026, China}
\author{Wei Yi}
\email{wyiz@ustc.edu.cn}
\affiliation{Laboratory of Quantum Information, University of Science and Technology of China, Hefei 230026, China}
\affiliation{Anhui Province Key Laboratory of Quantum Network, University of Science and Technology of China, Hefei 230026, China}
\affiliation{CAS Center for Excellence in Quantum Information and Quantum Physics, Hefei 230026, China}
\affiliation{Hefei National Laboratory, University of Science and Technology of China, Hefei 230088, China}

\begin{abstract}
In a strongly interacting Rydberg atom array, the dynamics are often constrained to the decoupled and fragmented Hilbert subspaces, representing an intriguing paradigm for nonergodicity.
Here we investigate the Hilbert-space fragmentation in the degenerate subspaces arising from the interplay between tunable detuning and arbitrary long-range interactions in a one-dimensional Rydberg-atom array with lattice-translation symmetry.
We analytically establish the conditions for the emergence of subspaces and their subsidiary fragmentations in the thermodynamic limit, enumerate the emergent kinetic constraints at the discrete dynamical points where dynamics are possible at all, and demonstrate the conditions for strong fragmentation and the rich scaling behaviors of the fragments.
Our results provide systematic and analytical understandings of Hilbert-space fragmentation in Rydberg arrays, and shed light on engineering nonergodic many-body dynamics beyond the PXP model.
\end{abstract}

\maketitle

{\it Introduction.---} 
Understanding the emergence of nonergodicity is fundamentally important for the study of thermalization (or the lack thereof) in isolated quantum many-body systems~\cite{Deutsch1991, Srednicki1994, Rigol2008, AdvancesinPhysics65292016, Basko2006,OganesyanHuse2007,Serbyn2013,Imbrie2016,  RevB821744112010, PhysRevB901742022014, CommunMathPhys33210172014, AnnuRevCondensMatterPhys6152015,RevModPhys910210012019}. Besides well-known mechanisms such as the disorder-induced many-body localization~\cite{Basko2006,OganesyanHuse2007,Serbyn2013,Imbrie2016, RevB821744112010, PhysRevB901742022014, CommunMathPhys33210172014, AnnuRevCondensMatterPhys6152015,RevModPhys910210012019}, it has become increasingly clear that constrained dynamics in clean systems can also give rise to ergodicity breaking~\cite{Moudgalya2022RPP,Turner2018, Choi2019, Serbyn2021, Nature5515792017, PhysRevLett1220406032019,PhysRevLett1261030022021, PhysRevB981551342018, PhysRevLett1221734012019,Mukherjee2021, PhysRevB1051251232022,PhysRevX100110472020,PhysRevB1041551172021,PhysRevB1100454182024,PhysRevLett1290906022022,Nature636802024}. A prominent example is the PXP model arising from Rydberg atom arrays~\cite{PhysRevLett1261030022021, PhysRevB981551342018, PhysRevLett1221734012019, PhysRevB1051251232022}, where the strong nearest-neighbor (NN) interactions impose a kinetic constraint in the form of the Rydberg blockade, yielding decoupled Hilbert subspaces and constrained oscillatory dynamics typical of quantum many-body scars~\cite{Turner2018,Choi2019,Serbyn2021,Moudgalya2022RPP}.

The high tunability of a Rydberg atom array also lends itself to explorations beyond the conventional paradigms~\cite{Browaeys2020, PhysRevB1111443132025,PhysRevResearch6L0420462024}.
For instance, by adjusting the detuning of the global Rydberg coupling, the system can be brought into the antiblockade regime~\cite{PhysRevB1130143172026,Ates2007PRL,Amthor2010PRL,Marcuzzi2017PRL}, with kinetic constraints and dynamics complementary to those under the Rydberg blockade. 
When longer-range interactions are considered, constraints associated with different interaction ranges are expected to collectively reshape the Hilbert-space structure, leading to potential Hilbert-space fragmentation~\cite{Moudgalya2022RPP,PhysRevLett1242076022020,PhysRevB1082051042023,PhysRevB1080451272023,PhysRevResearch50432392023,PhysRevX120110502022} and highly nonergodic dynamics.
While these insights suggest rich phenomena under the interplay of detuning and long-range interactions, a systematic understanding of their combined impact is still lacking.

In this work, we study the Hilbert-space structure and constrained dynamics of a one-dimensional strongly interacting Rydberg array with tunable detuning and interaction range. 
We show that Hilbert subspaces are well-defined (that is, protected by energy gaps) in the thermodynamic limit, if and only if the detuning is rational multiples of the interaction strength. More importantly, for each interaction range, for instance, NN or next-nearest-neighbor (NNN), a discrete set of detunings exists, under which dynamics within any subspace are possible at all. At these detunings, the subspaces are typically strongly fragmented, and the fragment dimensions can scale polynomially or exponentially with the system size, depending on the parameters. This in turn leads to constrained dynamics dictated by the dimension and connectivity of the corresponding fragment. 
Adopting an auxiliary-fermion description~\cite{PhysRevD980345052018,PRXQuantum30403172022,PhysRevResearch4L0320372022}, we show that the subspace fragmentation derives from emergent kinetic constraints, which manifest as local conserved quantities.

\begin{figure}
    \centering
    \includegraphics[width=\linewidth]{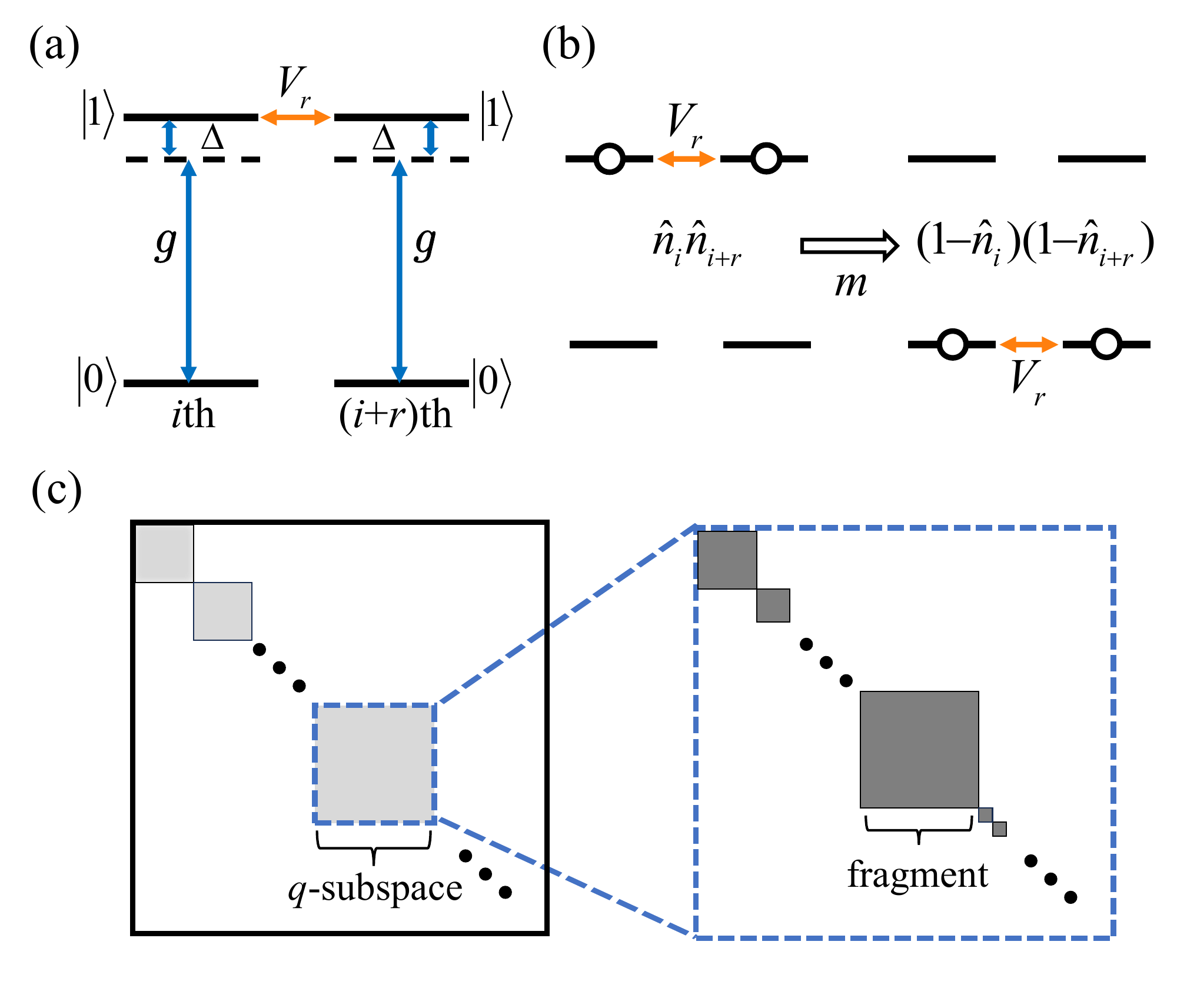}
    \caption{(a) Level scheme for atoms in a one-dimensional Rydberg array, where the ground states $|0\rangle$ are coupled to the Rydberg states $|1\rangle$. The atoms interact through the van der Waals interaction $V_r$, where $r$ denotes the interaction range. 
    (b) As the dimensionless parameter $m$ is tuned, the effective interaction continuously changes from the Rydberg-blockade type $\hat{n}_{i}\hat{n}_{i+r}$ to the antiblockade type $(1-\hat{n}_{i})(1-\hat{n}_{i+r})$. (c) Hierarchical structure of the Hilbert space. The full Hilbert space is partitioned into the $q$-subspaces (left), for appropriate values of $m$. Each $q$-subspace is further fragmented into multiple disconnected fragments (right), under the emergent kinetic constraints.}
    \label{fig1}
\end{figure}

{\it Model and Hilbert-space structure.---}
As illustrated in Fig.~\ref{fig1}(a), we consider a one-dimensional Rydberg atom array~\cite{ChinesePhysB300203052021,PhysRevLett1301830012023,
Bluvstein2021Science}, where atoms in the ground state $|0\rangle$ are coupled to the Rydberg state $|1\rangle$ with Rabi frequency $g$ and detuning $\Delta$. The atoms interact through a repulsive van der Waals interaction $V_r$, where the subscript $r\in \mathbb{Z}_+$ denotes the interaction range. The Hamiltonian of the system is then
\begin{align}
\hat{H}=g\sum_{i}\hat{X}_{i}+V\hat{Q}, 
\label{eq2}
\end{align}
where $\hat{Q}=\sum_{i}(\sum_{r}k_{r}\hat{n}_{i}{\hat{n}}_{i+r}-m\hat{Z}_{i})$, $V_r=k_rV$ ($r\geq 1$) with $V$ the interaction strength, $\hat n_i=\ket{1}_i\bra{1}$ is the local projector onto the Rydberg state, and $m=\Delta/V$ is the ratio between the detuning and interaction strength. 
Note that, while $k_r$ denotes the decay of interaction strength over range, we fix $k_1=1$ throughout this work.
The Pauli operators $\hat{X}_i$ and $\hat{Z}_i$ of the $i$th site are defined through $\hat{X}_i=|1\rangle_i\langle 0|+|0\rangle_i\langle 1|$, and $\hat{Z}_i=|1\rangle_i\langle 1|-|0\rangle_i\langle 0|$, respectively.

Let us first examine the case of strong NN interactions, with $V\gg g$ and $k_r=0$ for $r>1$. The detuning-modified dimensionless interaction becomes
\begin{align}
\hat{Q}=\sum_{i}(m-\hat{n}_{i})(m-\hat{n}_{i+1})+mL(1-m),
\label{eq3}
\end{align}
where $L$ is the system size. In the limit $m=0$, the dynamics are effectively described by the PXP model~\cite{NatRevPhys65662024}.
Therein, the strong NN interactions dominate, and prohibit NN Rydberg excitations under the Rydberg blockade. Dynamics are thus restricted to the degenerate eigen-subspaces of the interaction term $\hat{Q}$, dubbed the $q$-subspaces ($q$ being the eigenvalue of $\hat{Q}$).
In contrast, by setting $m=1$, the system enters the complementary antiblockade regime [see Fig.~\ref{fig1}(b)], where the process of flipping two NN Rydberg states into the ground states (or, equivalently, flipping two NN spins from spin-up to spin-down states) is prohibited, and dynamics are again restricted to the $q$-subspaces. 
Naively, these observations suggest similar scenarios should hold as $m$ continuously varies, connecting the blockade and antiblockade regimes. However, as we show below, in the thermodynamic limit, the $q$-subspaces are only well-defined when $m$ is rational. And the $q$-subspaces are strongly fragmented in general [see Fig.~\ref{fig1}(c)], yielding constrained dynamics within these subspace fragments.

\begin{figure}
    \centering
   \includegraphics[width=\linewidth]{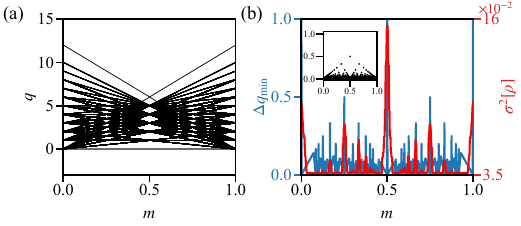}
\caption{(a) Eigenspectrum of $\hat{Q}$ as a function of $m$ with only NN interactions.
(b) Minimum spectral gap $\Delta q_{\min}$ (blue) and variance of the density of states
$\mathrm{\sigma^{2}}[\rho]$ (red) versus $m$.
 We evaluate the minimum spectral gap according to $\Delta q_{\min}=\min_{i\neq j}|q_i-q_j|$, where $\{q_i\}$ denotes the set of distinct eigenvalues of $\hat Q$. For the density of states, we divide the entire range of the eigenvalues $\{q_i\}$ into
$N_{\rm bin}$ equally-spaced intervals, and the variance of the density of states is calculated through
$\sigma^2[\rho]=N_{\rm bin}^{-1}\sum_{j=1}^{N_{\rm bin}}
[\rho_j-\overline{\rho}]^2$,
with
$\overline{\rho}=N_{\rm bin}^{-1}\sum_{j=1}^{N_{\rm bin}}\rho_j$. Here, 
$\rho_j$ represents the density of states for the $j$th interval.
The inset shows the Thomae function. We take $L=12$ with a periodic boundary condition for numerical calculations.}
 \label{fig2}
\end{figure}

{\it Subspace and dynamical points.---}
In Fig.~\ref{fig2}(a), we show the eigenspectrum of $\hat{Q}$ for the NN-interaction-only case with varying $m$. 
Notably, the spectral gaps do not appear to be well-resolved throughout. Indeed, we establish that the minimum spectral gap $\Delta q_{\text{min}}$ here is analytically given
by~\cite{supp}
\begin{equation}
    \lim_{L\to \infty}\Delta q_{\text{min}} = \text{Thomae}(2m),
\end{equation}
where the Thomae function is defined as~\cite{BartleSherbert2011}
\begin{equation}
 \text{Thomae}(x) =\begin{cases} 
0, &\text{for}\,\, x \notin \mathbb{Q}, \\
\dfrac{1}{p}, & \text{for}\,\, x= \dfrac{p'}{p},\ \gcd(p',p)=1,
\end{cases}
\end{equation}
with $p'\in \mathbb Z,\;p \in \mathbb{N}_{+}$.
For more general finite-range interactions with $k_r\neq 0$ ($r>1$), $\Delta q_{\text{min}}$ is also directly related to the Thomae function, such that, in the thermodynamic limit,  $\hat{Q}$ only acquires a finite  minimum spectral gap when $m$ is a rational number. Hence, the existence of $q$-subspaces is conditional. 

The analytical results above are numerically confirmed in Fig.~\ref{fig2}(b),
where, with varying $m$, the minimum gap (blue) and the variance of the density of states $\sigma^2[\rho]$ (red) both show features consistent with that of the Thomae function (inset).

\begin{figure*}[t]
\centering

\begin{minipage}[t]{0.34\textwidth}
    \vspace{0pt}
    \centering
\includegraphics[width=\linewidth]{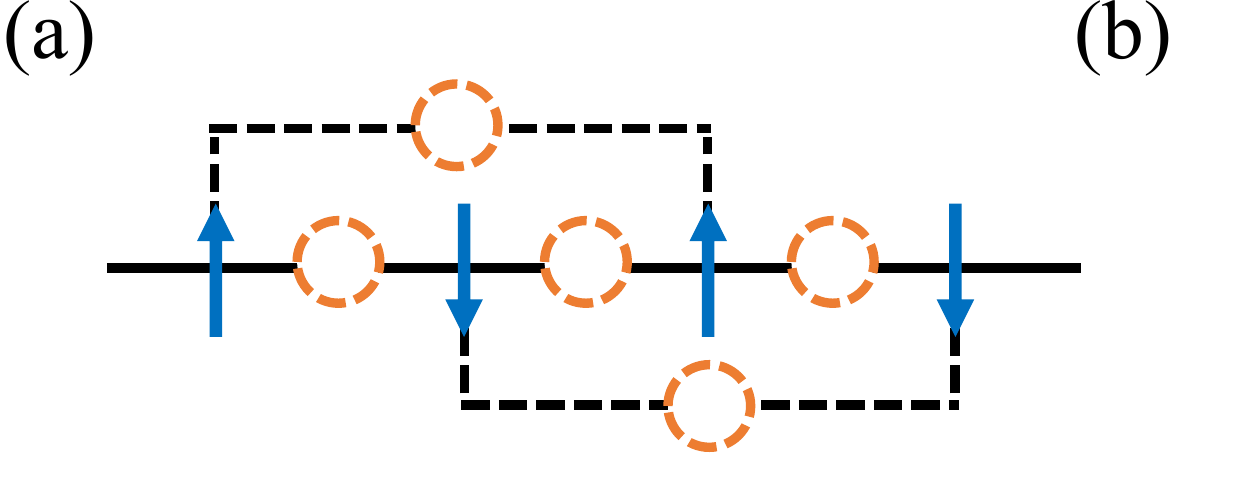}\\
\end{minipage}
\hfill
\begin{minipage}[t]{0.6\textwidth}
    \vspace{0.2cm}
    \raggedright
    \hspace*{-3.8em}
    \begin{tabular}{cccc}
        \toprule
        $m_r$ & $\hat{A}_{i,r}$ & $\hat{F}_{i,r}$ & local conserved quantity\\
        \midrule
        $0$
        & $\hat{n}_{i}\hat{n}_{i+r}$
        & $\hat{f}_{i,i-r}\hat{f}_{i,i+r}$
        & $2\hat{f}_{i,i+r}^{\dagger}\hat{f}_{i,i+r}+\hat{Z}_{i}+\hat{Z}_{i+r}=0,4$\\

        $k_{r}/2$
        & $\hat{Z}_{i}\hat{Z}_{i+r}/4$
        & $(\hat{f}^{\dagger}_{i,i-r}\hat{f}_{i,i+r}+\hat{f}^\dagger_{i,i+r}\hat{f}_{i,i-r})$
        & $2\hat{f}^{\dagger}_{i,i+r}\hat{f}_{i,i+r}+\hat{Z}_{i}\hat{Z}_{i+r}-1=0$ \\

        $k_{r}$
        & $(1-\hat n_i)(1-\hat n_{i+r})$
        & $\hat{f}^\dagger_{i,i-r}\hat{f}^\dagger_{i,i+r}$
        & $2\hat{f}^{\dagger}_{i,i+r}\hat{f}_{i,i+r}-\hat{Z}_{i}-\hat{Z}_{i+r}=0,4$ \\
        \bottomrule
    \end{tabular}
\end{minipage}

\caption{(a) Schematics of the spin-fermion representation, for a Rydberg chain with NN and NNN  interactions. Blue arrows denote onsite spins, and auxiliary fermions (dashed circles) reside on NN (solid black) and NNN (dashed black) bonds. 
(b) Tabulated correspondence between $m_r$, effective interaction $\hat{A}_{i,r}$, contribution of auxiliary fermions to the spin-fermion interaction $\hat{F}_{i,r}$, and the local conserved quantities.
}
\label{fig3}
\end{figure*}

When the $q$-subspaces are well-defined, the spectral gap of $\hat H$ diverges as $V/g\to\infty$. The dynamics are therefore restricted to each $q$-subspace, and governed by the corresponding projected Hamiltonian
\begin{equation}\label{PHP}
    \hat{H}_{\mathrm{eff}}^{(q)} = \hat{P}_q \left(  \sum_i \hat{X}_i \right) \hat{P}_q,
\end{equation}
where $\hat{P}_{q}$ projects onto a given $q$-subspace. 
Importantly, we find that dynamics within a subspace are only possible when~\cite{supp}
\begin{equation}
 m\in \mathcal{M}:=\left \{\sum_{r}m_{r}\ \bigg|\ m_{r}\in\{0,k_{r}/2,k_{r}\}\right \},
 \label{eq6}
\end{equation}
where $\hat{H}_{\mathrm{eff}}^{(q)}\neq 0$. Each existing interaction range thus contributes a discrete set of $m$ according to Eq.~(\ref{eq6}), which are henceforth referred to as the dynamical points.

Analytically, the dimensionless interaction at these dynamical points can be written as
\begin{equation}
\hat{Q}=\sum_{i,r}\left[k_{r}\hat{A}_{i,r}+m_{r}(k_{r}-m_{r})/k_{r}\right],
\label{eqQ}
\end{equation}
where $\hat{A}_{i,r}=\hat n_i \hat n_{i+r}$ for $m_r = 0$, $\hat{A}_{i,r}=\hat Z_i \hat Z_{i+r}/4$ for $m_r = k_r/2$, and $\hat{A}_{i,r} =(1-\hat n_i)(1- \hat n_{i+r})$ for $m_r = k_r$. 
When only NN interactions are present, $\hat{Q}$ describes blockade-, Ising-, and antiblockade-type interactions, respectively, at these dynamical points. When multiple interaction ranges coexist, the interplay of the long-range interactions $\hat A_{i,r}$ leads to rich subspace fragmentation and dynamics. 

{\it Fragmentation under emergent kinetic constraints.---}
A key feature of $q$-subspaces at the dynamical points is that the projected Hamiltonian $\hat{H}^{(q)}_{\text{eff}}$ acquires emergent kinetic constraints, leading to further fragmentation and constrained dynamics.

Fragmentation of a $q$-subspace manifests as its decomposition into
disconnected Krylov subspaces [see Fig.~\ref{fig1}(c)].
For a product state $\ket{\psi}$, the associated Krylov subspace is
$\mathcal{K}(\ket{\psi})=\operatorname{span}\{\ket{\psi},
\hat{H}_{\mathrm{eff}}^{(q)}\ket{\psi},
[\hat{H}_{\mathrm{eff}}^{(q)}]^2\ket{\psi},\ldots\}$, which not only establishes the connectivity between different product states, but also constitutes a fragment of the corresponding $q$-subspace.
In particular, when $\hat{H}_{\mathrm{eff}}^{(q)}\ket{\psi}=0$, based on Eq.~(\ref{PHP}), 
$\hat{P}_q\hat{X}_i\hat{P}_q\ket{\psi}=0$ for all $i$, and
$\ket{\psi}$ forms an isolated fragment with
$\dim\mathcal{K}(\ket{\psi})=1$.
Otherwise, at least one local spin flip (at site $i$ for instance) satisfies
$\hat{P}_q\hat{X}_i\hat{P}_q\ket{\psi}\neq0$. These local spin flips connect product states within the same $q$-subspace, and
Krylov subspaces (or fragments) can be systematically constructed by consecutive applications of local spin flips $\hat{P}_q \hat{X}_i \hat{P}_q$ on all possible product states.
Such a strategy applies for arbitrary system size $L$~\cite{supp}.
These fragments generally exhibit different dynamics depending on their sizes, with more strongly constrained dynamics for smaller fragments.

To gain physical insight of the underlying kinetic constraints for the subspace fragmentation, we introduce auxiliary fermions living on the bonds of long-range interactions [see Fig.~\ref{fig3}(a)], such that the constraint imposed by the projector in Eq.~(\ref{PHP}) is recast into local conserved quantities enforced by the Pauli exclusion principle~\cite{supp,NatRevPhys65662024}. 
Note that an equivalent auxiliary hard-core boson representation was adopted in Ref.~\cite{PhysRevB1070943022023} to study the ground-state phase transitions. Here, by contrast, we focus on the global Hilbert-space structures.
Crucially, for a given interaction range $r$, different effective interactions arise at different dynamical points, yielding distinct spin-fermion interactions.

At the dynamical points with $m_r=0$, the effective interaction term in Eq.~(\ref{eqQ}) takes the form $ \hat{n}_i\hat{n}_{i+r}$, thereby imposing a kinetic constraint consistent with the Rydberg blockade. Specifically, in the spin-fermion representation, the auxiliary fermions live on bonds connecting spins with the same orientation. A local spin flip $\hat S_i^\dagger=|1\rangle_i\langle 0|$ is accompanied by the annihilation of fermions on the two neighboring bonds, a process given by $\hat S_i^\dagger \hat F_{i,r}$ with $\hat F_{i,r}=\hat f_{i,i-r}\hat f_{i,i+r}$. Here $\hat{f}_{i,j}$ is the fermion annihilation operator for the bond between the $i$th and $j$th spins.
Thus, the Pauli exclusion principle prevents spins on the two ends of a bond to simultaneously flip to the spin-up state. 
Equivalently, an emergent kinetic constraint is imposed in the form of the local conserved quantity $2\hat{f}_{i,i+r}^{\dagger}\hat{f}_{i,i+r}+\hat{Z}_{i}+\hat{Z}_{i+r}=0\,\,\text{or}\,\,4$, 
where $0$ and $4$ correspond to unblocked and blocked configurations, respectively~\cite{supp}. 

When $m_r=k_{r}/2$, the effective interaction in Eq.~(\ref{eqQ}) becomes $\hat{Z}_i\hat{Z}_{i+r}/4$. In this case, the auxiliary fermions live on bonds connecting spins with opposite orientations,
and a local spin flip is equivalent to the hopping of a fermion across the flipped spin. 
Specifically, the spin flip $\hat S_i^\dagger$ is accompanied by the hopping process $\hat F_{i,r}=\hat f_{i,i-r}^{\dagger}\hat f_{i,i+r}+\hat f_{i,i+r}^{\dagger}\hat f_{i,i-r}$, while the local conserved quantity becomes $2\hat{f}^{\dagger}_{i,i+r}\hat{f}_{i,i+r}+\hat{Z}_{i}\hat{Z}_{i+r}-1=0$. 

Finally, when $m_r=k_r$, the effective interaction in Eq.~(\ref{eqQ}) becomes $(1-\hat{n}_i)(1-\hat{n}_{i+r})$, which follows directly from the $m_r=0$ case under the particle-hole transformation $\hat{n}_{i}\to 1-\hat{n}_{i}$. The corresponding spin-fermion interaction and local conserved quantity can be similarly obtained~\cite{supp}.

In summary, different interaction ranges correspond to different auxiliary-fermion models with distinct local conserved quantities, as tabulated in Fig.~\ref{fig3}(b).
Since the kinetic constraints for different interaction ranges are independent, the subspace Hamiltonian~(\ref{PHP}) in the spin-fermion representation takes the form
\begin{align}
\hat{H}_{\text{S-F}}=\sum_{i}\hat{S}^{\dagger}_{i}\left(\prod_{r}\hat{F}_{i,r}\right)+\text{H.c.}.
\label{eq9}
\end{align} 
As longer range interactions are included, more local conserved quantities emerge, leading to increasingly strong constraints on the subspace structure. Thus, subspace fragmentation generically arises under the emergent kinetic constraints, leading to highly nonergodic dynamics.

{\it Examples.---}
We consider two examples, both having NN and NNN interactions with $k_2<k_1$. 
Note that $k_2$ is tunable in practice through a zig-zag spatial reconfiguration of the Rydberg atom array~\cite{PhysRevLett1350934032025}.
In the case of $m=0$, the system corresponds to the well-known PPXPP model, whose constrained dynamics~\cite{PhysRevB1031043022021,PhysRevB1052451372022,PhysRevLett1341604012025} and ground-state phase transitions~\cite{PhysRevB1070943022023} in the $q=0$ subspace have been extensively studied.
Here we focus on the other two types of dynamical points in the examples below. 

\begin{figure}
    \centering   \includegraphics[width=\linewidth]{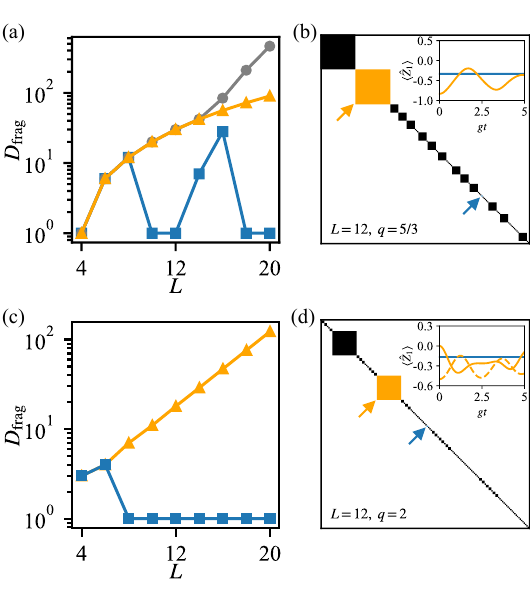}
\caption{(a)(b) Fragmentation in the $q=k_{2}(L-2)/2$ subspace with $m=k_{2}/2$. 
In (a), the dimensions of the largest fragment (gray) and the one containing the single-excitation state $\ket{\psi}=\ket{1000\cdots0}$ (orange) are polynomial in the system size $L$ for large $L$. 
While that of the smallest fragment (blue) remains finite.
Detailed fragment structure in the subspace is shown in (b) for $L=12$. Typical dynamics of  $\langle \hat{Z}_{1}\rangle$ within the orange and blue fragment (indicated by arrows) are illustrated in the inset.
(c)(d) Fragmentation in the $q=k_{2}L/2$ subspace with $m=k_{2}$.
In (c), the dimension of the fragment containing the N\'eel state (orange) is exponential in the system size $L$. The fragment is also the largest one within the subspace. 
That of the smallest fragment (blue) remains finite.
Detailed fragment structure in the subspace is shown in (d) for $L=12$. Typical dynamics of  $\langle \hat{Z}_{1}\rangle$ within the orange and blue fragments (indicated by arrows) are illustrated in the inset.
Two different initial states are chosen for dynamics in the orange fragment~\cite{supp}. While the solid curve (inset) represents more extensive dynamics within the fragment, the dashed one shows the characteristic periodic oscillations associated with many-body scars.
For numerical calculations, we take $k_{2}=1/3$ and $V/g=1000$. }
 \label{fig4}
\end{figure}

First, we consider the case with $m=k_{2}/2$. 
Adopting the auxiliary-fermion approach and eliminating the spin degrees of freedom~\cite{supp}, the subspace Hamiltonian is mapped onto an integrable fermionic model~\cite{SciPostPhysCore40102021,SciPostPhys100992021}
\begin{align}
\hat{H}_{\text{eff}}^{(q)} = \sum_{l} \hat{c}^{\dagger}_{l-1} \left( 1 - \hat{c}_{l}^{\dagger}\hat{c}_{l} \right) \hat{c}_{l+1} + \mathrm{H.c.},
\end{align}
with $\hat{c}_{l}=\hat{f}_{l,l+2}$.

A central question here is how the fragment dimension $D_{\mathrm{frag}}$ scales with the system size $L$. 
To see this, we start with a fixed product state, for instance, the single-excitation state $\ket{\psi}=\ket{1000\cdots0}$, which lies within the $q=k_2(L-2)/2$ subspace. 
We then numerically evaluate the relevant fragment dimension by constructing the Krylov subspace to which it belongs.
According to the auxiliary-fermion representation, there are no fermions on even lattice sites, and the dynamics are restricted to the odd ones. The Hamiltonian within the fragment is then given by
\begin{equation}
    \hat{H}_{\text{frag}} = \sum_{l\in \mathrm{odd}} \hat{c}^{\dagger}_{l} \hat{c}_{l+2} + \mathrm{H.c.},
    \label{eqT-c}
\end{equation}
which implies that the dimension of the fragment scales as $D_\text{frag}\propto\binom{L/2}{2}$,
polynomially in $L$. This is numerically confirmed in Fig.~\ref{fig4}(a), where
we also show the largest and smallest fragment dimensions in the relevant $q$-subspace for different system sizes. 
While the largest fragment dimension grows polynomially in $L$ for sufficiently large $L$,  the smallest fragments remain finite, indicating that strictly finite-dimensional fragments persist in the thermodynamic limit. Dynamics thus remain nonergodic even within the $q$-subspace.

To further illustrate the fragmentation structure and the resulting dynamics, we focus on the $L=12$ case, where the relevant subspace is $q=5k_{2}$ (for the single-excitation state). Within this subspace, fragments of different sizes exhibit markedly different dynamics. 
As shown in Fig.~\ref{fig4}(b), 
while intra-fragment dynamics survive within larger fragments (inset orange), they are frozen within the smallest fragment (inset blue).

We now turn to the second example with $m=k_{2}$.
Herein, the NN term enforces the Rydberg blockade, and the NNN term imposes the antiblockade condition. 
We focus on the N\'eel state $\ket{1010\dots}$, which resides in the subspace with $q=k_{2}L/2$.
In the spin-fermion representation, auxiliary fermions fully occupy the NNN bonds connecting even-site spins, thus enforcing spin-down states on even sites.
Consequently, the dynamics are confined to odd lattice sites, where an allowed spin flip on an odd site $i$ annihilates auxiliary fermions on the adjacent NNN bonds with $\hat{f}_{i,i-2}$ and $\hat{f}_{i,i+2}$. Therefore, in the atomic representation, the Hamiltonian within the corresponding fragment is
\begin{equation}\label{HfragPXP}
\hat{H}_{\text{frag}} = \sum_{i\in \text{odd}}\hat{n}_{i-2}\hat{X}_{i}\hat{n}_{i+2}.
\end{equation}
This is essentially the PXP model, meaning the dimension of the fragment should grow exponentially with $L$.

The pictures above are confirmed in Fig.~\ref{fig4}(c), where we also find that the fragment containing the N\'eel state happens to be the largest one. Similar to the previous example, the dimension of the smallest fragments remains finite for sufficiently large $L$. Numerical results for the fragment distribution and the corresponding dynamics within the same $q=k_{2}L/2$ subspace are shown in Fig.~\ref{fig4}(d),
where the fragment marked in orange corresponds precisely to the PXP model of Eq.~(\ref{HfragPXP}). 
We also show representative fragment-constrained dynamics in the inset. 
While the dashed orange curve demonstrates the periodic oscillations characteristic of the many-body scars of the PXP model, the blue curve indicates completely frozen dynamics within the smallest fragment.

Although the two examples discussed above exhibit qualitatively different scaling behaviors of fragment dimensions, both display strong Hilbert-space fragmentation~\cite{Moudgalya2022RPP}, characterized by the vanishing relative weight of the largest fragment in the thermodynamic limit.
Indeed, one can show analytically that strong fragmentation necessarily emerges at the dynamical points
if at least one $m_r$ exists such that $m_r=0$ or $m_r=k_r$~\cite{supp}.

{\it Discussion.---}
We have clarified the general Hilbert-space structure and fragmentation mechanism in Rydberg arrays under the interplay of tunable detuning and arbitrary-range interactions. 
While the Thomae-function profile of the minimum gap can be probed using the many-body Ramsey interferometry~\cite{supp,Zeiher2016}, our analytical results provide useful guide for experimental implementation and control of the many-body dynamics, either by fine tuning parameters and initial states, or by engineering kinetic constraints.
Our study can also serve as the starting point for more complicated scenarios with broken translational symmetry, where the Hilbert-space structure and the kinetic constraints are modified, for instance, by local defects or disorder~\cite{supp,PhysRevB1122051032025}.

This work is supported by the National Natural Science Foundation of China (Grant No. 12374479), and Quantum Science and Technology-National Science and Technology Major Project (Grant No. 2021ZD0301904).

{\it Data availability.---}The data supporting this study are available in Ref.~\cite{ZenodoData}.

% ========================== Supplemental Material ==========================
\clearpage
\onecolumngrid
\phantomsection
\label{sec:supplement}
\addcontentsline{toc}{section}{Supplemental Material}

\begin{center}
{\large\bfseries Supplemental Material for ``Emergent Kinetic Constraints and Subspace Fragmentation in Rydberg Arrays''\par}
\end{center}

\setcounter{section}{0}
\setcounter{equation}{0}
\setcounter{figure}{0}
\setcounter{table}{0}

\renewcommand{\thesection}{S\arabic{section}}
\renewcommand{\theequation}{S\arabic{equation}}
\renewcommand{\thefigure}{S\arabic{figure}}
\renewcommand{\thetable}{S\arabic{table}}

% Distinct PDF destinations for the restarted counters.
\renewcommand{\theHsection}{supp.\arabic{section}}
\renewcommand{\theHequation}{supp.\arabic{equation}}
\renewcommand{\theHfigure}{supp.\arabic{figure}}
\renewcommand{\theHtable}{supp.\arabic{table}}

In this Supplemental Material, we provide details on the general connection of minimum spectral gap and Thomae function, the dynamical points, the spin-fermion representation, details of the Hilbert-space fragmentation, details of the initial states, the impact of single defect, and the experimental implication of our work.

\section{Thomae function in Rydberg arrays}\label{app:labelA}
In the main text, we have discussed the close connection between the Thomae function and the minimum spectral gap of the operator $\hat{Q}$. Here, we first prove Eq.~(3) in the main text, that the minimum spectral gap is exactly given by the Thomae function for systems with only nearest-neighbor (NN) interactions. We then extend the argument to general finite-range interactions, for which the minimum spectral gap is also related to the Thomae function.

\subsection{I. NN interactions only}
We first focus on the NN case with $k_1=1$. 
Since $[\hat Q,\hat Z_i]=0$ for all positions $i$, $\hat Q$ is diagonal in the product states
$|\mathbf n\rangle=|n_1,\ldots,n_L\rangle$, with $n_i\in\{0,1\}$ in the basis of $\hat{Z}_i$. 
The corresponding eigenvalue is
\begin{equation}
q(\mathbf n)=\sum_{i}n_i n_{i+1}-2m\sum_i n_i+mL.
\end{equation}
Defining the total excitation number $N=\sum_i n_i\in\mathbb Z$ and the blockade number $B=\sum_i n_i n_{i+1}\in\mathbb Z$, one sees that, up to a constant shift, the spectrum depends only on $(N,B)$, with \begin{equation} q=B-2mN . \end{equation}
Therefore, the spectral difference between any two product-state configurations is
\begin{equation}
\Delta q=\Delta B-2m\,\Delta N .
\end{equation}
For a chain of infinite length, at a fixed total excitation number $N$, the blockade number $B$ can take all integer values $B=0,1,\dots,N-1$. In the thermodynamic limit, $N$ ranges over all natural numbers. Therefore, for arbitrary integers $Z_1,Z_2\in\mathbb{Z}$, one can always find two product-state configurations $(N_1,B_1)$ and $(N_2,B_2)$ such that
\begin{equation}
\Delta B = B_2 - B_1 = Z_1,\quad 
\Delta N = N_2 - N_1 = Z_2 .
\end{equation}

We first consider the case in which $2m$ is rational. Let
\begin{equation}
2m=\frac{p'}{p},
\end{equation}
where $ p'\in \mathbb Z,\;p\in\mathbb N_+, \; \gcd(p,p')=1$. Then the spectral difference between any two configurations satisfies
\begin{equation}
\Delta q=\Delta B-\frac{p'}{p}\Delta N ,
\end{equation}
and thus
\begin{equation}
p\,\Delta q=p\Delta B-p'\Delta N\in\mathbb Z .
\end{equation}
It follows that, if $\Delta q\neq0$, one must have $|\Delta q|\ge 1/p$, implying that the level spacing cannot be smaller than $1/p$. On the other hand, by B\'ezout's lemma~\cite{HardyWright2008}, for co-prime integers $p$ and $p'$, there exist integers $x,y$ such that $px-p'y=1$. Using the result above, one can choose two configurations satisfying $\Delta B=x$ and $\Delta N=y$, yielding
\begin{equation}
\Delta q=x-\frac{p'}{p}y=\frac{1}{p}.
\end{equation}
Hence, in the thermodynamic limit, the minimum nonzero level spacing is ${1}/{p}$.

If $2m$ is irrational, then by the Diophantine approximation~\cite{HardyWright2008}, for any $\varepsilon>0$, there exist integers $n\ge1$ and $k\in\mathbb Z$ such that $|k-2mn|<\varepsilon$. Therefore, in the thermodynamic limit, the infimum of the spectral spacing is zero, that is, $\lim_{L\to \infty}\Delta q_{\text{min}}(2m)=0$.

In summary, under NN-only interactions, the minimum positive spectral gap between degenerate eigenspaces of $\hat Q$ varies with detuning $2m$ as
\begin{equation}
\lim_{L\to \infty}\Delta q_{\text{min}}=
\begin{cases}
0, & 2m\notin\mathbb Q,\\[6pt]
\dfrac{1}{p}, & 2m=\dfrac{p'}{p},\ \gcd(p',p)=1 ,
\end{cases}
\end{equation}
which is precisely the Thomae function.

\subsection{ II. General finite-range interactions}

We now consider general finite-range interactions of the form
\begin{equation}
q=\sum_{r} k_r B_r -2mN+mL,
\end{equation}
where
$
N=\sum_i n_i,\, B_r=\sum_i n_i n_{i+r}\quad (1\le r\le R),
$
and $R$ is the maximal interaction range. Here only finitely many $k_r$ are nonzero, with each $k_r\in\mathbb{Q}$.
Therefore, the spectral difference between any two configurations is
\begin{equation}
\Delta q=\sum_{r=1}^{R} k_r \Delta B_r-2m\Delta N.
\end{equation}
To characterize all possible spectral differences, we need to determine all possible vectors of the form
$(\Delta N,\Delta B_1,\dots,\Delta B_R)$. 
Although the system is defined on an infinite chain, for any configuration containing only finitely many excitations, 
$N$ and all $B_r$ depend only on a finite region containing those excitations. 
Such a configuration can therefore be represented by a finitely supported Boolean vector, obtained by trimming away the zero tails on both sides
\begin{equation}
    \ket{\mathbf n}
=
|\underbrace{\cdots 000}_{\text{discarded}}
\,
\overbrace{1\cdots 1}^{\vec n}
\,
\underbrace{000 \cdots}_{\text{discarded}}
\rangle, 
\end{equation}
where $\vec n$ is the Boolean vector supported on the minimal interval containing all excitations. 
The vectors $\vec n_1,\vec n_2,\dots$ below are all understood in this sense, with zeros padded on both sides when they are embedded back into the infinite chain.
This representation involves no loss of generality, because in the thermodynamic limit, different local excitation patterns can always be separated far enough that their contributions to $N$ and $B_r$ remain independent.
For any finitely supported Boolean vector $\vec n$, define the map
\begin{equation}
\Phi(\vec n)=(N,B_1,B_2,\dots,B_R)\in \mathbb N^{R+1}.
\end{equation}
We are interested in the set of all possible differences
\begin{equation}
\mathcal D=\left\{\Phi(\vec n_1)-\Phi(\vec n_2)\mid \vec n_1,\vec n_2 \in \mathcal{B}\right\},
\end{equation}
where $\mathcal{B}$ is the set of finitely supported Boolean vectors. Clearly, $\mathcal{D}\subset \mathbb Z^{R+1}$. We now show that in fact $\mathcal D=\mathbb Z^{R+1}$.

First, let $\vec n$ and $\vec n'$ be two finitely supported Boolean vectors. If we concatenate them with an all-zero interval of length at least $R$ in between (denoted by $ {\vec 0^{(R)}}$), then there are no cross-term contributions to any $B_r$ ($1\le r\le R$), since the maximal interaction range is $R$. Therefore,
\begin{equation}\label{Phiplus}
\Phi(\vec n\oplus {\vec 0^{(R)}}\oplus \vec n')=\Phi(\vec n)+\Phi(\vec n').
\end{equation}
Here, $\oplus$ denotes the direct sum of Boolean vectors. Using this additivity, one readily checks that $\mathcal D$ forms an additive subgroup. The zero vector belongs to $\mathcal D$ since $\Phi(\vec n)-\Phi(\vec n)=0$ for any $\vec n$. If $d=\Phi(\vec n_1)-\Phi(\vec n_2)\in\mathcal D$, then exchanging $\vec n_1$ and $\vec n_2$ gives $-d\in\mathcal D$, so $\mathcal D$ is closed under inverses. Finally, for
$d_1=\Phi(\vec n_1)-\Phi(\vec n_2)$ and $d_2=\Phi(\vec n_3)-\Phi(\vec n_4)$,
Eq.~(\ref{Phiplus}) implies
\begin{align}
d_1+d_2
&=\Phi(\vec n_1)-\Phi(\vec n_2)+\Phi(\vec n_3)-\Phi(\vec n_4) \nonumber\\
&=\Phi(\vec n_1\oplus {\vec 0^{(R)}}\oplus \vec n_3)-\Phi(\vec n_2\oplus {\vec 0^{(R)}}\oplus \vec n_4)\in\mathcal D.
\end{align}
Thus $\mathcal D$ is an additive subgroup of $\mathbb Z^{R+1}$.

We next show that $\mathcal D$ contains all standard basis vectors, and hence generates the whole lattice $\mathbb Z^{R+1}$. Define
\begin{equation}
e_0=(1,0,0,\dots,0),\quad e_r=(0,\dots,0,1,0,\dots,0),
\end{equation}
where, for $1\le r\le R$, the vector $e_r$ has its $(r+1)$th component equal to $1$. To construct $e_0$, we take the vacuum configuration $\vec 0 =(0,0,\cdots,0)$ and a configuration $\vec n^{(1)}$ containing only one excitation, which leads to $\Phi(\vec n^{(1)})-\Phi(\vec 0)=e_0$. 
To construct \(e_r\), we first consider an auxiliary vector obtained from a two-excitation configuration. For any \(1\le r\le R\), we take a configuration \(\vec n^{(2)}_r\) with exactly two excitations separated by distance \(r\). Then
\begin{equation}
\Phi(\vec n^{(2)}_r)-\Phi(\vec 0)
=(2,0,\dots,0,1,0,\dots,0)=:\tilde e_r,
\end{equation}
where the \((r+1)\)th component is \(1\). Since \(e_0,\tilde e_r\in\mathcal D\) and \(\mathcal D\) is an additive group, it follows that
$e_r=\tilde e_r-2e_0\in\mathcal D$.
Hence all basis vectors $e_0,e_1,\dots,e_R$ belong to $\mathcal D$, yielding $\mathcal D=\mathbb Z^{R+1}$. Equivalently, in the thermodynamic limit, for any integer vector $(Z_0,Z_1,\dots,Z_R)\in\mathbb Z^{R+1}$, one can always find two configurations such that
\begin{equation}
(\Delta N,\Delta B_1,\dots,\Delta B_R)=(Z_0,Z_1,\dots,Z_R).
\end{equation}
The spectral difference can therefore be written as
\begin{equation}
\Delta q=\sum_{r=1}^{R} k_r Z_r-2mZ_0.
\end{equation}
Now assume $k_r\in\mathbb Q$. Since $\{k_r\}_{r=1}^{R}$ is a finite set of rational numbers, the additive group they generate has a unique smallest positive element $\lambda$, namely, $\{\sum_{r} k_r Z_r\mid Z_r\in\mathbb Z\}=\lambda\mathbb Z$.
Therefore $\Delta q=\lambda Z-2mZ_0$, where $Z,Z_0\in\mathbb Z$. This is exactly analogous to the NN case, and thus, in the thermodynamic limit, the minimum nonzero spectral gap is
\begin{equation}
\lim_{L\to \infty}\Delta q_{\min}=\lambda\cdot \text{Thomae}
\left(\frac{2m}{\lambda}\right).
\end{equation}
Importantly, for general finite-range interactions with rational couplings $k_r$, the minimum gap in the spectrum of $\hat Q$ is still controlled by the Thomae function.

\section{Dynamical points}\label{app:labelB}
In this section, we prove that subspace dynamics are also possible at the dynamical points
\begin{equation}
 m\in \left \{\sum_{r}m_{r}\ \bigg|\ m_{r}\in\{0,k_{r}/2,k_{r}\}\right \}.
\end{equation}

If $\sum_i \hat{P}_q \hat{X}_i \hat{P}_q \neq 0$, there must exist $j,\alpha,\beta$ such that $\bra{q,\alpha} \hat{X}_j \ket{q,\beta} \neq 0$. Equivalently, there exists a spin flip $\hat X_{j}$, and a product state $\ket{q,\beta}$ such that $\hat{X}_j\ket{q,\beta} $ is still in the $q$-subspace. Here, $\ket{q,\alpha}$ and $\ket{q,\beta}$ denote different product states in the degenerate $q$-subspace, labeled by $\alpha$ and $\beta$ respectively. 

For an arbitrary product state $\ket{n_1 n_2\cdots n_L}$ with $n_i \in \{0,1\}$, we have
\begin{equation}
q = \sum_{i,r} k_r n_i n_{i+r} - 2m \sum_i n_i +\text{const}.
\end{equation}
When the spin at site $j$ is flipped, that is, $n_{j}\to1-n_{j}$, the change in $q$ is 
\begin{align}
\Delta q&=\sum_{r}k_{r}(1-2n_{j})(n_{j-r}+n_{j+r})-2m(1-2n_{j})\nonumber\\
&=(1-2n_{j})\left[\sum_{r}k_{r}(n_{j-r}+n_{j+r})-2m\right].
\end{align}
Imposing the condition $\Delta q=0$ yields
\begin{align}
m=\sum_r m_r:=\sum_{r}\frac{k_{r}}{2}(n_{j-r}+n_{j+r}),
\label{eqady}
\end{align}
where $(n_{j-r}+n_{j+r})=0,1,2$. Hence, for each interaction range $r$, $
m_{r}$ can take values $0,k_r/2$ and $k_r$.
Therefore, for a general finite-range interaction with maximal range $R$, there are $3^R$ dynamical points. Substituting a given set $\{m_r\}$ back into the Hamiltonian $\hat{H}=g\sum_{i} \hat{X}_{i}+V\hat{Q}$, one obtains
\begin{equation}
    \hat{H}=\sum_{i}g\hat X_{i}+\sum_{i,r}Vk_{r}\hat{A}_{i,r}+\text{const},
\end{equation}
where
\begin{equation}
\hat{A}_{i,r}
=
\begin{cases}
\hat n_i \hat n_{i+r},\;&m_{r}=0,\\
\hat Z_i \hat Z_{i+r}/4,\;&m_{r}=k_r/2,\\
(1-\hat n_i)(1-\hat n_{i+r}),\;&m_{r}=k_r.
\end{cases}
\end{equation}
\begin{figure}
    \centering
   \includegraphics[width=0.6\linewidth]{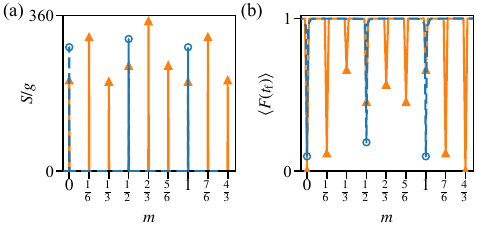}
\caption{
(a) Sum of the Frobenius norms $S$ as a function of $m$ for $L=12$. 
(b) Mean fidelity as a function of $m$. Parameters are $L=8$, $V/g=1000$, and $t_{\mathrm f}g=1.5$.
In (a) and (b), blue and orange correspond to the cases $(k_1,k_2)=(1,0)$ and $(1,1/3)$, respectively, and the markers denote the dynamical points predicted by Eq.~(\ref{eqady}).
}
    \label{figdynaimical}
\end{figure}

In Fig.~\ref{figdynaimical}(a)(b), we numerically confirm the condition for dynamical points
using two examples, one with only NN interactions, one with both NN and NNN interactions. For this purpose, we calculate Frobenius norm of the projected Hamiltonian 
\begin{equation}
    S=\sum_q \| \hat{H}^{(q)}_{\text{eff}} \|_{\mathrm F},
\end{equation}
where $\|\hat{A}\|_{\mathrm F}=\sqrt{\mathrm{Tr} (\hat{A}^\dagger \hat{A})}$, to determine whether $\hat H^{(q)}_{\text{eff}}$ is nonzero.
In both cases, $S$ is nonzero only at the dynamical points [Fig.~\ref{figdynaimical}(a)], confirming $\hat{H}_{\mathrm{eff}}^{(q)}\neq 0$. 
More relevantly, the presence of dynamical points can be demonstrated using real-time evolutions. We initialize the system in the product states $\{\ket{\psi_i}\}$ from a $q$-subspace, and evaluate the mean fidelity
\begin{equation}
\langle F(t_{\mathrm f})\rangle=\sum_{i=1}^{N_q}
|\bra{\psi_i} e^{-i\hat{H}t_{\mathrm f}} \ket{\psi_i}|^2/N_q,
\end{equation}
at the final time $t_\text{f}$. Here, $N_q$ is the total number of the product states in the $q$-subspace.
As shown in Fig.~\ref{figdynaimical}(b), the fidelity deviates appreciably from $1$ only near the dynamical points, consistent with the conclusion above. 

\section{Auxiliary fermions and local conserved quantities}\label{app:labelC}

\subsection{I. Local Conserved Quantities and Constrained Spin Flips}
\begin{figure*}
    \centering
    \includegraphics[width=0.9\textwidth]{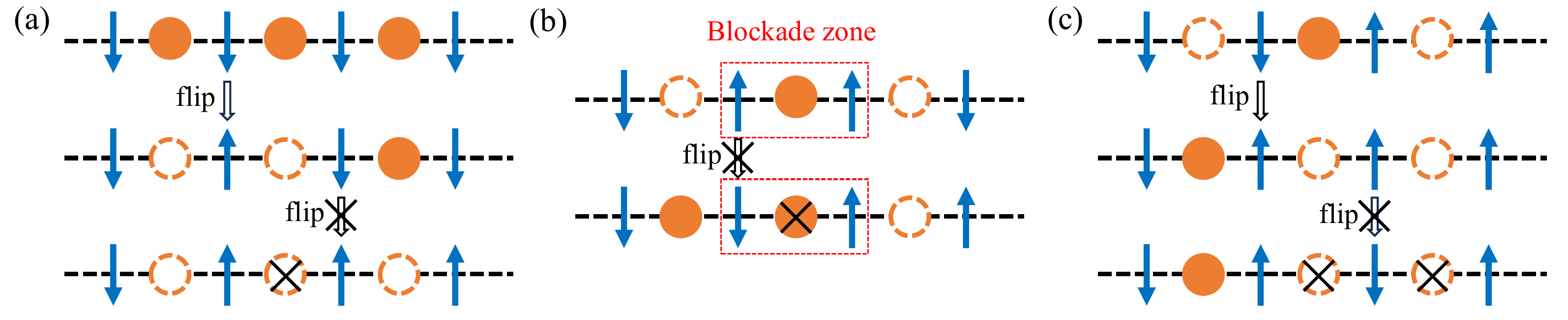}
    \caption{
Schematic illustration of the spin-fermion representation and the corresponding constrained local spin-flip processes. Filled circles denote the presence of an auxiliary fermion, while empty circles denote its absence. Blue arrows denote spins. 
(a) For $\hat Q=\sum_{i}\hat n_i\hat n_{i+1}$, an allowed spin flip annihilates the fermions on the two adjacent bonds, which forbids a subsequent flip of a neighboring spin. 
(b) For $\hat Q=\sum_{i}\hat n_i\hat n_{i+1}$, the auxiliary fermion within the blockade zone forbids the spin flip.
(c) For $\hat Q=\sum_{i}\hat Z_i\hat Z_{i+1}/4$, the fermions live on domain walls, and an allowed spin flip is accompanied by the hopping of the fermion.
}
    \label{fig:blockadezone}
\end{figure*}

In the main text, we have briefly discussed the local conserved quantities that arise after auxiliary fermions are introduced. In this section, we elaborate on this picture in more details. Using different types of NN interactions as intuitive examples, we explain how these local conserved quantities emerge and how they determine the allowed spin-flip processes. The same framework can be directly applied to general finite-range interactions.

For the Hamiltonian $\hat H = g\sum_i \hat X_i + V\hat Q$, we consider the case with the dimensionless interaction $\hat Q = \sum_{i}\hat n_i \hat n_{i+1}$. In the limit $V/g \to \infty$, depending on the initial product state, the dynamics is restricted to subspaces with fixed eigenvalue $q$ of $\hat Q$. Therefore, only spin-flip processes with $\Delta q=0$ are allowed. For $\hat Q=\sum_{i}\hat n_i\hat n_{i+1}$, $\Delta q=0$ means that a spin can flip only when both of its neighboring spins are in the spin-down state. 
The constraints between spins can be realized by introducing additional degrees of freedom, in the form of auxiliary fermions, on the bonds connecting the interacting spins. 
In this spin-fermion representation, a local spin flip from down to up is accompanied by the annihilation of the auxiliary fermions on the two adjacent bonds, that is, $ \hat{S}_i^{\dagger}\hat{f}_{i,i-1}\hat{f}_{i,i+1} $, such that the constraints between neighboring spins are effectively enforced by the Pauli exclusion principle. Following this construction, to obtain a representation fully equivalent to the original spin model, we need to require that a fermion occupies the bond between two neighboring spins when both of them are in the spin-down state. As illustrated in Fig.~\ref{fig:blockadezone}(a), this immediately forbids flipping two neighboring spins to the spin-up state, since such a process would require annihilating the same fermion twice and is therefore excluded by the Pauli principle. In this case, the fermion and its two adjacent spins satisfy the local constraint
\begin{equation}
2\hat f_{i,i+1}^{\dagger}\hat f_{i,i+1}+\hat Z_i+\hat Z_{i+1}=0.
\label{C1}
\end{equation}
On the other hand, for product states containing a local configuration with adjacent spin-up states, which we call a blockade zone, the corresponding local interaction satisfies $n_i n_{i+1}=1$. In this case, flipping either of the two spins must be completely forbidden to preserve the local interaction $n_i n_{i+1}$. This is achieved by fixing a fermion on bond within the blockade zone, as illustrated in Fig.~\ref{fig:blockadezone}(b). The associated local constraint then reads
\begin{equation}
2\hat f_{i,i+1}^{\dagger}\hat f_{i,i+1}+\hat Z_i+\hat Z_{i+1}=4.
\label{C2}
\end{equation}

The above construction can be generalized straightforwardly to \(\hat Q=\sum_i (1-\hat n_i)(1-\hat n_{i+1})\). Since this interaction is related to \(\hat Q=\sum_i \hat n_i \hat n_{i+1}\) by the particle-hole transformation \(\hat Z_i\to -\hat Z_i\), the corresponding local conserved quantities can be obtained directly by substituting this transformation into Eqs.~(\ref{C1}) and (\ref{C2}).

For $\hat Q=\sum_i \hat Z_i \hat Z_{i+1}/4$, the condition $\Delta q=0$ implies that a spin can flip only when its two neighboring spins are in opposite orientations. As in the previous case, auxiliary fermions can be introduced without changing the physical content. In this representation, the fermions live on domain walls, namely the bonds connecting two opposite spins. As shown in Fig.~\ref{fig:blockadezone}(c), each allowed spin flip is accompanied by the hopping of a fermion between the two adjacent bonds, described by $ \hat{S}_i^{\dagger} \left( \hat{f}_{i,i-1}^{\dagger}\hat{f}_{i,i+1} + \hat{f}_{i,i+1}^{\dagger}\hat{f}_{i,i-1} \right) $. The corresponding emergent local conserved quantity is \begin{equation}
    2\hat f_{i,i+1}^{\dagger}\hat f_{i,i+1}+\hat Z_i\hat Z_{i+1}-1=0.
\end{equation}
% \sout{We note that the introduction of auxiliary fermions enlarges the original Hilbert space and thereby introduces redundant degrees of freedom, which are commonly understood as an emergent gauge structure. In the present construction, such redundancy can be consistently removed by restricting to the physical subspace defined by the local conserved quantities. Specifically, these local constraints establish a correspondence between spin configurations and fermionic configurations, ensuring that the spin degrees of freedom can be faithfully encoded in terms of the auxiliary fermions. As a result, one can eliminate the redundant degrees of freedom and reformulate the problem entirely within a fermionic description, without altering the underlying physical content of the model.}

\subsection{
II. Mapping between the Spin Hilbert Space and the Spin--Fermion Hilbert Space}
In this section, we detail the mapping of the auxiliary fermion representation. We note from the outset that, while the mapping is exact in expressing the emergent kinetic constraints as the local conserved quantities, the kinetic constraints themselves are valid in the limit $V\gg g$, which is 
consistent with typical experimental conditions and serves as the general presumption for the study of Hilbert-space fragmentation in strongly interacting Rydberg atoms~\cite{PRXQuantum30403172022,PhysRevResearch4L0320372022,PhysRevB1070943022023}.

\begin{figure}
    \centering
    \includegraphics[width=0.55\linewidth]{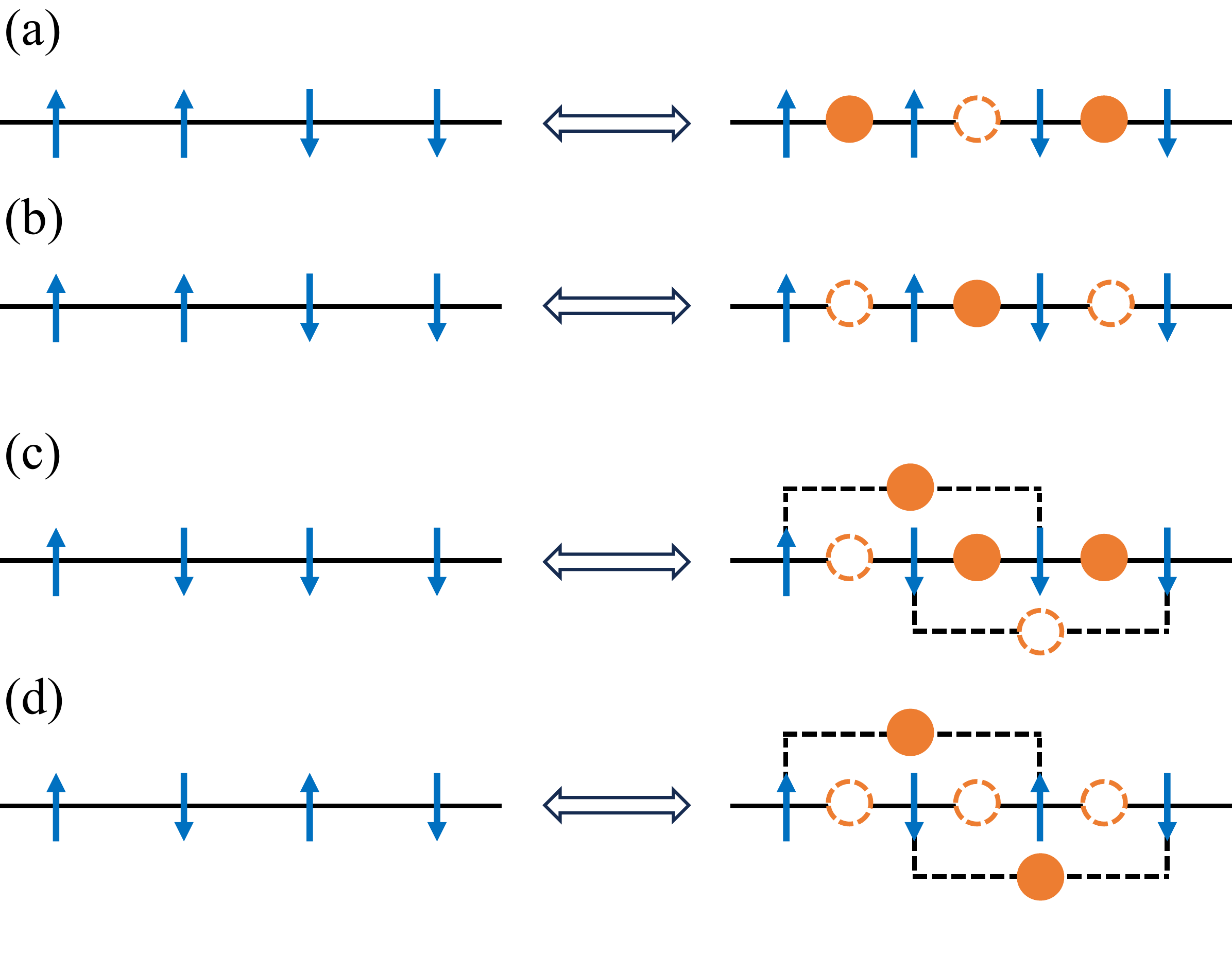}
    \caption{Spin--fermion mapping from \(\mathcal H_{\mathrm{s}}\) to the physical subspace of \(\mathcal H_{\mathrm{ext}}\). Blue arrows denote spin (atom) states, while filled (open) orange circles denote occupied (unoccupied) auxiliary-fermion modes. Solid and dashed links represent nearest- and next-nearest-neighbor bonds, respectively. (a) Nearest-neighbor blockade interaction for $m=m_{1}=0$. (b) Nearest-neighbor Ising interaction for $m=m_{1}=k_{1}/2$. (c) Nearest-neighbor blockade and next-nearest-neighbor Ising interactions for $m=m_{1}+m_{2},m_{1}=0,m_{2}=k_{2}/2$. (d) Blockade interactions at both ranges for $m=m_{1}+m_{2},m_{1}=0,m_{2}=0$.}
    \label{fig:mapping}
\end{figure}

For a system of $L$ atoms, the original Hilbert space of the atom array is
\begin{equation}
    \mathcal H_{\mathrm{s}}
    =
    \bigotimes_{i=1}^{L}\mathcal H_i^{\mathrm{s}},
\end{equation}
where the local Hilbert space of atom $i$ is
\begin{equation}
    \mathcal H_i^{\mathrm{s}}
    =
    \operatorname{span}
    \left\{
        \ket{0}_i,\ket{1}_i
    \right\}.
\end{equation}
Here, $\ket{0}_i$ and $\ket{1}_i$ denote the ground and Rydberg states, respectively. The corresponding product-state basis is
\begin{equation}
    \ket{\boldsymbol n}_{\mathrm{s}}
    =
    \bigotimes_{i=1}^{L}\ket{n_i}_i
    =
    \ket{n_1,n_2,\ldots,n_L}_{\mathrm{s}},
    \qquad
    n_i\in\{0,1\}.
\end{equation}
After introducing auxiliary fermions on the interaction bonds, the enlarged Hilbert space becomes
\begin{equation}
\mathcal H_{\mathrm{ext}}
=
\mathcal H_{\mathrm{s}}\otimes\mathcal H_{\mathrm{f}},
\end{equation}
where $\mathcal H_{\mathrm{f}}$ is the auxiliary-fermion Hilbert space. Although $\mathcal H_{\mathrm{ext}}$ is larger than the original Hilbert space, the local conserved quantities would select a physical sector with the same number of degrees of freedom as $\mathcal H_{\mathrm{s}}$. This sector is specified by the allowed eigenvalues of the local conserved quantities
\begin{equation}
\begin{cases}
2\hat f_{i,i+r}^{\dagger}\hat f_{i,i+r}
+\hat Z_i+\hat Z_{i+r}=0\ \text{or}\  4,
&
m_r=0,
\\[2mm]
2\hat f_{i,i+r}^{\dagger}\hat f_{i,i+r}
+\hat Z_i\hat Z_{i+r}-1=0,
&
m_r=k_r/2,
\\[2mm]
2\hat f_{i,i+r}^{\dagger}\hat f_{i,i+r}
-\hat Z_i-\hat Z_{i+r}=0\ \text{or} \ 4,
&
m_r=k_r,
\end{cases}
\label{eq:local_constraints_reply}
\end{equation}

% \begin{equation}
% |\boldsymbol n\rangle_{\mathrm s}
% \underset{}{\overset{}{\longleftrightarrow}}
% |\boldsymbol n\rangle_{\mathrm s}
% \otimes
% \bigotimes_{(i,i+r)}
% \left|
% n^{\text{f}}_{i,i+r}(\boldsymbol n)
% \right\rangle,
% \label{eq:isometry_reply}
% \end{equation}
% \begin{equation}
% \ket{\boldsymbol n}_{\mathrm s}
% \longleftrightarrow
% \ket{\boldsymbol n}_{\mathrm s}
% \otimes
% \bigotimes\nolimits_{(i,i+r)}
% \left|
% n_{i,i+r}^{\mathrm f}(\boldsymbol n)
% \right\rangle .
% \label{eq:isometry_reply}
% \end{equation}

Therefore, for any given on-site spin configuration, these local conditions uniquely determine the occupation of every auxiliary-fermion mode. The resulting one-to-one mapping is thus dictated by (\ref{eq:local_constraints_reply}) as
\begin{equation}
\ket{\boldsymbol n}_{\mathrm s}
\longleftrightarrow
\ket{\boldsymbol n}_{\mathrm s}
\otimes
\left|
n_{1,2}^{\mathrm f}\, ,n_{1,3}^{\mathrm f},\cdots,n_{1,1+R}^{\mathrm f}\;,n_{2,3}^{\mathrm f}\, ,n_{2,4}^{\mathrm f},\cdots,n_{2,2+R}^{\mathrm f}\;,n_{3,4}^{\mathrm f}\,, \cdots
\right\rangle,
\label{eq:isometry_reply}
\end{equation}
where $n_{i,i+r}^{\mathrm f}\in\{0,1\}$ denotes the occupation number of the auxiliary-fermion mode on the bond $(i,i+r)$. These occupation numbers specifies an auxiliary-fermion Fock state in the occupation-number representation. By construction, the mapped state defined in Eq.~(\ref{eq:isometry_reply}) is normalized and antisymmetric under the exchange of auxiliary fermions. Explicitly,

\begin{equation}
n^{\mathrm f}_{i,i+r}
=
\begin{cases}
\displaystyle
\frac{1+Z_i Z_{i+r}}{2},
&
m_r=0\ \text{or}\ m_r=k_r,
\\[8pt]
\displaystyle
\frac{1-Z_i Z_{i+r}}{2},
&
m_r=k_r/2,
\end{cases}
\label{eq:fermion_mapping_reply}
\end{equation} 
with
\begin{equation}
Z_i=2n_i-1\in\{-1,1\}.
\end{equation}
% \begin{equation}
% |\boldsymbol n\rangle_{\mathrm s}
% \underset{}{\overset{}{\longleftrightarrow}}
% |\boldsymbol n\rangle_{\mathrm s}
% \otimes
% \prod_{(i,i+r)}
% \left(
% \hat f^\dagger_{i,i+r}
% \right)^{n^{\mathrm f}_{i,i+r}(\boldsymbol n)}
% |0\rangle_{\mathrm f},
% \label{eq:isometry_reply}
% \end{equation}
% where the product runs over all interaction bonds $(i,i+r)$ in an arbitrary but fixed ordering, and $|0\rangle_{\mathrm f}$ denotes the Fock vacuum of the auxiliary fermions. Different choices of the ordering amount only to different phase conventions for the fermionic Fock basis and do not affect the occupation mapping. Here, \(n^{\mathrm f}_{i,i+r}(\boldsymbol n)\) is the fermion occupation
% number on the bond \((i,i+r)\), given by

Thus, in the blockade ($m_r=0$) and antiblockade ($m_r=k_r$) cases, auxiliary fermion modes are occupied on bonds connecting parallel spins, whereas in the Ising case ($m_r=k_r/2$), the modes are occupied for bonds connecting antiparallel spins.
These cases are schematically illustrated in Fig.~\ref{fig:mapping}.

Importantly, since the local constraints uniquely determine the auxiliary-fermion configurations, the mapping is exactly one-to-one between the original Hilbert space and the physical subspace of \(\mathcal H_{\mathrm{ext}}\) selected by the local conserved quantities. Based on this one-to-one mapping, the Pauli exclusion principle effectively replaces the role of the original subspace projector. It then follows that the action of the original projected Hamiltonian $\hat{H}^{(q)}_{\mathrm{eff}}$ on a spin-space state in $\mathcal{H}_s$ is equivalent to the action of the spin--fermion Hamiltonian $\hat{H}_{\text{S-F}}$ on the corresponding state in the spin--fermion Hilbert space $\mathcal{H}_{\mathrm{ext}}$. 

\section{More details of the Hilbert-space fragmentation}\label{app:labelD}
\subsection{I. Flippable local configurations and connectivity}

As stated in the main text, if
$
\hat{H}_{\mathrm{eff}}^{(q)}\ket{\psi}\neq 0,
$
there exists at least one site $i$, such that
$
\hat{P}_{q}\hat{X}_{i}\hat{P}_{q}\ket{\psi}\neq 0.
$
This means that the spin at site $i$ is flippable within the $q$-subspace. In this section, we make this condition explicit in terms of local product-state configurations, and show how these local spin-flip rules determine the connectivity between product states.

Consider an arbitrary product state
\begin{equation}
\ket{\psi}=\ket{\cdots n_{i-R}\cdots n_{i-1}n_i n_{i+1}\cdots n_{i+R}\cdots},
\end{equation}
with $n_\ell\in\{0,1\}$, in a given $q$-subspace. A spin flip at site $i$ is allowed only if it does not change the eigenvalue $q$, that is, only if the action of $\hat X_i$ keeps the state inside the same $q$-subspace. This requirement can be formulated as a condition on the pairs of spins symmetrically located around site $i$. Concretely, for each interaction range $r=1,2,\dots,R$, the local configuration must satisfy
\begin{equation}
\left(n_{i-r},n_{i+r}\right)=
\begin{cases}
(0,0), & m_r=0, \\[4pt]
(0,1)\ \text{or}\ (1,0), & m_r=k_r/2, \\[4pt]
(1,1), & m_r=k_r,
\end{cases}
\label{eq:local_flip_rule_app}
\end{equation}
so that a spin flip at site $i$ leaves $q$ unchanged. Whenever Eq.~\eqref{eq:local_flip_rule_app} is satisfied for all relevant $r$, the state $\ket{\psi}$ is connected by $\hat{H}_{\mathrm{eff}}^{(q)}$ to another product state that differs only at site $i$,
\begin{equation}
\hat X_i\ket{\psi}=\ket{\cdots n_{i-R}\cdots n_{i-1}\,\bar{n}_i\,n_{i+1}\cdots n_{i+R}\cdots},
\end{equation}
where $\bar{n}_i=1-n_i$.
In this way, each product state can be viewed as a vertex in a connectivity graph of the $q$-subspace, and an edge is drawn whenever two product states differ by one allowed local spin flip. Starting from an arbitrary product state $\ket{\psi}$, one may determine all flippable sites according to Eq.~\eqref{eq:local_flip_rule_app}, generate all directly connected states $\{\hat X_i\ket{\psi}\}$, and then repeat the same procedure for each newly obtained state. Iterating this construction produces the full set of product states that are connected to $\ket{\psi}$, yielding the fragment containing $\ket{\psi}$.

It is important to emphasize that the ability of spin flip of different sites is not independent. Once a spin at site $i$ is flipped, the local neighborhood is modified, and may either enable or suppress subsequent flips at other sites, thereby connecting different product states into fragments of different sizes and structures.

\subsection{II. Strong Fragmentation in Representative Multirange Subspaces}
In the main text, we have established that, for the NNN model, both the $m=k_{2}/2$ case in the $q=k_{2}(L-2)/2$ subspace and the $m=k_{2}$ case in the $q=k_{2}L/2$ subspace exhibit strong subspace fragmentation, in the sense that the largest fragment occupies a vanishing fraction of the corresponding $q$-subspace in the thermodynamic limit. Here the precise definition for such a strong fragmentation is~\cite{Moudgalya2022RPP}
\begin{equation}
\lim_{L\to\infty}D^{(q)}_{\text{max}}/D_q=0,
\end{equation}
where $L$ is the system size, $D^{(q)}_{\text{max}}$ is the dimension of the largest fragment and $D_q$ is the dimension of the corresponding $q$-subspace. The vanishing relative weight of the largest fragment establishes that no single fragment becomes dominant in the thermodynamic limit, which is the defining feature of strong fragmentation.
In this section, we present, as examples, two additional subspaces $q=k_{2}(L-4)/2$ for $m=k_{2}/2$ and $q=k_{2}(L/2-2)+2k_{1}$ for $m=k_{2}$. These subspaces also exhibit subspace fragmentation. Together with the scaling results for the two representative subspaces discussed in the main text, they show that our conclusions are general.

\begin{figure}
    \centering
   \includegraphics[width=0.5\linewidth]{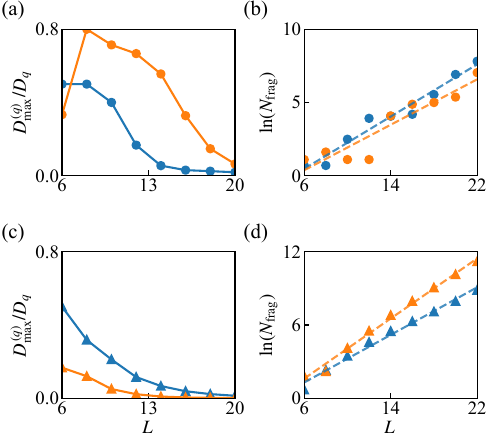}
\caption{(a)(b) Ratio $D^{(q)}_{\text{max}}/D_q$ and the scaling of the logarithm of the number of fragments, $\ln N_{\mathrm{frag}}$, as a function of the system size for $m=k_{2}/2$. Blue and orange circles denote the $q=k_{2}(L-2)/2$ and $q=k_{2}(L-4)/2$ subspaces, respectively. Dashed lines in (b) indicate linear fits. 
(c)(d) Ratio $D^{(q)}_{\text{max}}/D_q$ and the scaling of the logarithm of $N_{\mathrm{frag}}$, $\ln N_{\mathrm{frag}}$, as a function of the system size for $m=k_{2}$. Blue and orange triangles denote the $q=k_{2}L/2$ and $q=k_{2}(L/2-2)+2k_{1}$ subspaces, respectively. Dashed lines in (d) indicate linear fits. We fix $k_{2}=1/3$ for all calculations.}
    \label{strongfragment}
\end{figure}

For $m=k_{2}/2$, we consider the fragment generated from the product state $\ket{1001000\dots000}$ in the $q=k_{2}(L-4)/2$ subspace. This fragment admits a fermionic interpretation in which two fermions occupy the odd sites and two fermions occupy the even sites, with their motion described by the density-dependent hopping Hamiltonian
$
\hat{H}_\text{frag}
=
\sum_{l}
\hat{c}^{\dagger}_{l-1}
\left(1-\hat{n}^{f}_{l}\right)
\hat{c}_{l+1}
+\mathrm{H.c.}.
$
For $m=k_{2}$, we focus on the fragment generated from a product state $|11101010\cdots\rangle$ obtained from the N\'eel state by flipping a single spin-down state to spin-up. This state lies in the subspace $q=k_{2}(L/2-2)+2k_{1}$. The corresponding fragment is described by the effective Hamiltonian
$\hat{H}_{\text{frag}}=
\sum_{\substack{i\in \mathrm{odd} \\ i\notin \{1,3\}}}
\hat{n}_{i-2}\hat{X}_{i}\hat{n}_{i+2},$
which corresponds to a constrained dynamics on the odd subchain, with two spins frozen by the NN blockade.

We now present the numerical results supporting the conclusion of strong fragmentation for the four subspaces considered here. 
As shown in Fig.~\ref{strongfragment}(a), for $m=k_{2}/2$, the ratio $D^{(q)}_{\text{max}}/D_q$ decreases with the system size in both the $q=k_{2}(L-2)/2$ and $q=k_{2}(L-4)/2$ subspaces. Fig.~\ref{strongfragment}(c) shows that the same conclusion holds for the $q=k_{2}L/2$ and $q=k_{2}(L/2-2)+2k_{1}$ subspaces with $m=k_{2}$. Therefore, in all four cases, the largest fragment occupies a vanishing fraction of the corresponding $q$-subspace in the thermodynamic limit. This provides direct evidence for strong subspace fragmentation. Here, the essential criterion for strong fragmentation is not whether the largest fragment grows polynomially or exponentially in its absolute dimension, but whether it becomes dominant relative to the full $q$-subspace. Our results show that this does not occur in any of the four subspaces considered.

The exponential growth of the total number of disconnected fragments with the system size is also a common feature of strongly fragmented regimes. We therefore examine how the number of fragments scales with $L$.
Fig.~\ref{strongfragment}(b) shows how the number of disconnected fragments scales with $L$ for two subspaces at $m=k_2/2$, namely $q=k_{2}(L-2)/2$ and $q=k_{2}(L-4)/2$, shown as blue and orange circles, respectively. Fig.~\ref{strongfragment}(d) shows the corresponding results for the two $m=k_{2}$ subspaces, $q=k_{2}L/2$ and $q=k_{2}(L/2-2)+2k_{1}$, corresponding to the blue and orange triangles in Fig.~\ref{strongfragment}(c). In all four cases, the data for $\ln N_{\mathrm{frag}}$ are well captured by linear fits, demonstrating that the number of disconnected fragments increases exponentially with the system size. These results further show that the fragmented structure not only survives in the thermodynamic limit, but also proliferates rapidly with increasing $L$.

\subsection{III. Strong Fragmentation Induced by Blockade Configuration}

% In this section, we establish a general condition for strong fragmentation. If $m_r=0$ or $m_r=k_r$ for at least one interaction range $r$, the corresponding blockade configurations divide the relevant $q$-subspaces into disconnected fragments. We show that the relative dimension of the largest fragment vanishes in the thermodynamic limit, thereby establishing strong fragmentation.
In this section, we establish a sufficient condition for strong fragmentation. Specifically, if \(m_r=0\) or \(m_r=k_r\) for at least one interaction range \(r\), there necessarily exist \(q\)-subspaces that exhibit strong fragmentation. 
%We identify the structures in these \(q\)-subspaces (referred to below as blockade bonds) and show how these structures necessarily give rise to strong fragmentation.

\begin{figure}
    \centering
    \includegraphics[width=0.4\linewidth]{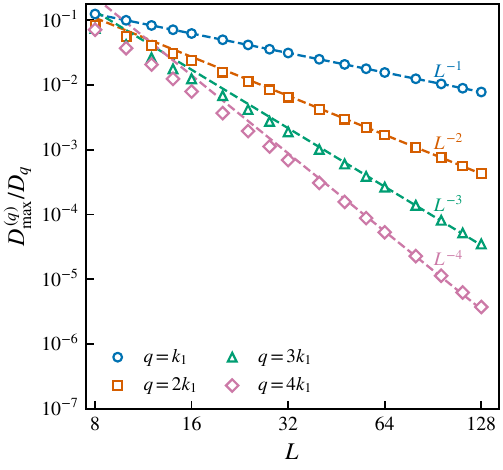}
\caption{Scaling of the ratio $D_{\max}^{(q)}/D_q$ with respect to the system size $L$, for $q=k_{1},2k_{1},3k_{1},4k_{1}$ with $k_{1}=1$ at $m=0$. 
Dashed lines indicate the exact scalings of $L^{-1}$, $L^{-2}$, $L^{-3}$, and $L^{-4}$, respectively.}
    \label{fig:strong fragmentaion}
\end{figure}

As an example, we first consider the simplest case with only nearest-neighbor interactions at $m=m_1=0$, where $ q=k_1\sum_i n_i n_{i+1}$. Each pair of adjacent Rydberg excitations (dubbed a blockade bond) contributes to \(q\), so that the total number of blockade bonds, \(B_1=\sum_i n_i n_{i+1}\), directly determines \(q\). 
In the following, we consider \(q\)-subspaces associated with configurations containing a nonzero number of \(B_1\) blockade bonds, and show that these subspaces are necessarily strongly fragmented. 
Each of these subspaces splits into dynamically disconnected fragments, where the dimension of the largest fragment scales as $D_{\max}^{(q)}/D_q\propto L^{-B_1}\longrightarrow0$ in the thermodynamic limit, as illustrated in Fig.~\ref{fig:strong fragmentaion} for $q=B_1k_{1}=k_{1},2k_{1},3k_{1},4k_{1}$. 
Here $D_q$ is the total dimension of the $q$-subspace, and $L$ is the system size. This scaling originates from the spatial configuration of the blockade bonds. A local spin flip cannot create, remove, or relocate a blockade bond without changing $q$. Configurations with 
the $B_1$ blockade bonds at different locations therefore belong to dynamically disconnected fragments. For any finite $B_1$, the number of possible arrangements of the $B_1$ blockade bonds scales as
\begin{equation}
\binom{L}{B_1}
\sim
\frac{L^{B_1}}{B_1!},
\end{equation}
which remains valid in the thermodynamic limit. The Hilbert-space dimension of the fragments associated with these configurations have the same leading-order scaling with $L$
\begin{equation}
\frac{D_{\max}^{(q)}}{D_q}
\propto
L^{-B_1}.
\end{equation}
Note that the special $q=0$ subspace is the Hilbert space of the conventional PXP model, which contains no blockade bonds and is not further fragmented.
The case of $m_1=k_1$ is related to that of $m_1=0$ by a particle-hole transformation, and exhibits the same fragmentation behavior. 
Since configurations with blockade bonds necessarily exist in the Hilbert space for any given set of parameters, strong fragmentation also necessarily exists in the case of $m_1=0$ or $m_1=k_1$.

More generally, if at least one $m_r$ exists such that $m_r=0$ or $m_r=k_r$, strong fragmentation is also necessarily present. 
For a given $r$ (assuming the presence of interactions at range $r$), we start from the case with $m_r=0$ (in the decomposition of $m$), 
and consider the Rydberg-blockade configuration over sites $i$ and $i+r$, 
meaning the Rydberg states are occupied on sites $i$ and $i+r$ (hence a range-$r$ blockade bond).
In this case, a local spin flip at either site is forbidden within the same $q$-subspace, as it
changes the value of $q$. Consequently, the total number of range-$r$ blockade bonds,
$
B_r\equiv\sum_i n_i n_{i+r},
$
is conserved during the dynamics. 

We now focus on configurations containing \(B_r\) blockade bonds of range \(r\) ($B_r\neq 0$), and examine how the corresponding \(q\)-subspace scales with the system size \(L\). For our proof here, it is sufficient to consider configurations with $B_r\neq 0$ but remains finite as $L\rightarrow \infty$.
Importantly, blockade bonds of the same range but at different spatial locations belong to distinct fragments of the same $q$-subspace, as they are not dynamically connected.
Assuming the lattice-translation symmetry, translating a configuration (which should contain at least one range-$r$ blockade bond) by an arbitrary number of lattice sites would end up in the same $q$-subspace.

However, translating the configuration also shifts the positions of the blockade bonds, so these translated configurations must belong to distinct fragments with the same dimension (by translation symmetry).
Applying this argument to the largest fragment gives
\begin{equation}
\frac{D_{\max}^{(q)}}{D_q}
\leq
\frac{1}{L},
\end{equation}
which vanishes as $L\rightarrow\infty$, thus establishing strong fragmentation.

\section{Initial states used in the numerical simulations}

In this section, we tabulate the initial states for the numerical simulations in Fig.~4 of the main text. 
The labels ``orange'' and ``blue'' refer to the corresponding curves shown in the insets of Fig.~4(b) and Fig.~4(d).

\begin{table}[ht]
\centering
\begin{tabular}{c@{\hspace{1.0cm}}l@{\hspace{1.0cm}}c}
\hline\hline
Figure & Curve & Initial state \\
\hline
Fig.~4(b) & orange  & $\ket{100000000000}$ \\
Fig.~4(b) & blue    & $\ket{010001000110}$ \\
\hline
Fig.~4(d) & orange solid   & $\ket{101010101010}$ \\
Fig.~4(d) & orange dashed  & $\ket{100010001000}$ \\
Fig.~4(d) & blue solid     & $\ket{001100100101}$ \\
\hline\hline
\end{tabular}
\caption{
Initial states for simulations in the insets of Fig.~4(b) and Fig.~4(d) of the main text.
}
\label{tab:initial_states_fig4}
\end{table}

\section{Impact of a Single Geometric Defect}

\begin{figure}
    \centering
\includegraphics[width=0.9\linewidth]{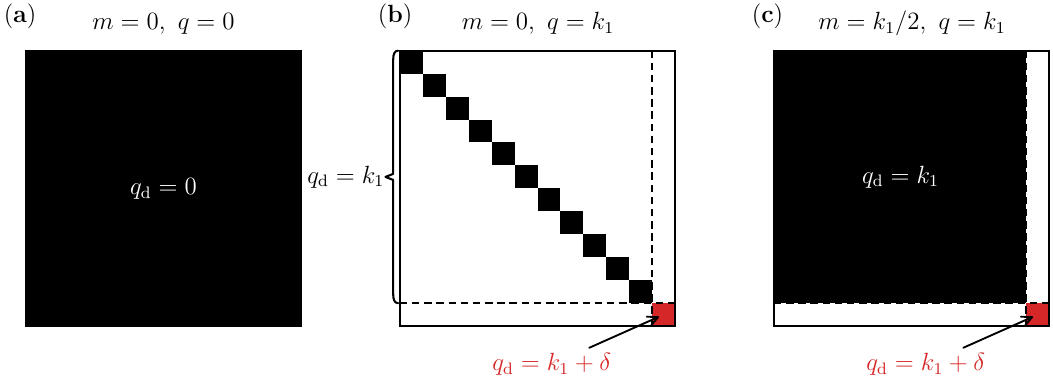}
    \caption{Impact of a single geometric defect on the fragmentation structure of a nearest-neighbor Rydberg chain with $L=12$. 
In each panel, the horizontal and vertical axes enumerate all product states in the corresponding $q$-subspace of the translation-invariant system, and the colored blocks represent fragments that are dynamically disconnected from the rest of the Hilbert space.
Red blocks denote fragments whose eigenvalues are shifted relative to those in the absence of the defect. The dashed lines indicate the defect-induced spectral splitting into different $q_{\mathrm d}$ sectors. 
(a) For $m=0$, the $q=0$ subspace neither splits spectrally nor changes its fragmentation structure. 
(b) For $m=0$, the $q=k_1$ subspace splits into the $q_{\mathrm d}=k_1$ and $q_{\mathrm d}=k_1+\delta$ sectors, while the connectivity of the original fragments remains unchanged. 
(c) For $m=k_1/2$, the $q=k_1$ subspace splits into the $q_{\mathrm d}=k_1$ and $q_{\mathrm d}=k_1+\delta$ sectors, and its connectivity is reduced from one fully connected fragment to two disconnected fragments. }
    \label{fig:defect_q_subspace}
\end{figure}
In this section, we consider a single geometric defect introduced by locally changing one atomic spacing and hence the Rydberg interaction strengths~\cite{PhysRevB1122051032025}. We study how this defect affects the gap-protected subspaces and their Hilbert-space connectivity in the strong-interaction limit.

In general, a single geometric defect can have three effects. First, it can lift 
degeneracies within a $q$-subspace and generate energy-level 
splittings. In the strong-interaction limit, any finite splitting of the 
interaction-term eigenvalues is amplified by the interaction strength $V$, leading to 
a finer structure of gap-protected subspaces. 
Second, at the dynamical points, the defect can modify the 
kinetic constraints and hence reshape the Hilbert-space connectivity. 
Third, when the system is not at a dynamical point, the defect can at most induce local dynamics with small connectivity, out of the previously frozen dynamics. 
In the following, we focus on the impact of defect at the dynamical points, as these cases involve larger connectivity and thus more interesting.
To summarize, we find that the impact of defect depends on the type of effective interaction: the connectivity generated by blockade-type interactions is robust against the local defect, whereas Ising- and antiblockade-type interactions generally acquire modified local spin-flip rules.

We introduce the single geometric defect by changing the spacing between atoms $\ell$ and $\ell+1$.
In the subsequently modified Hamiltonian~(1), the interaction term remains diagonal in the product-state basis, with eigenstates $|n_1n_2\cdots\rangle$ specified by $n_i=0,1$. The corresponding eigenvalue of the dimensionless interaction operator $\hat Q$ is then modified to
\begin{equation}
q_{\mathrm d}=\sum_{r=1}^{R}q_{d,r},
\label{eq:qd_general}
\end{equation}
where
\begin{equation}
q_{d,r}
=
\sum_j
\left(k_r+\delta_{j,j+r}\right)n_jn_{j+r}
-
m_r\sum_j Z_j .
\label{eq:qdr_general}
\end{equation}
Here, $r$ labels the interaction range, and $\delta_{j,j+r}$ denotes the defect-induced modification of the interaction strength on the bond $(j,j+r)$. 
% \revadd{For a range-$r$ bond $(j,j+r)$, we set $\delta_{j,j+r}=\delta_r$ if the bond crosses the modified spacing $(\ell,\ell+1)$, and $\delta_{j,j+r}=0$ otherwise. Here, $\delta_r$ is the corresponding change in the interaction strength.}
Equations~(\ref{eq:qd_general}) and (\ref{eq:qdr_general}) show that product states sharing the same eigenvalue $q$ in the absence of defect (with $\delta_{j,j+r}=0$), can acquire different eigenvalues after the defect is introduced. In the process, a $q$-subspace can further split into several defect-modified subspaces, which become dynamically disconnected in the strong-interaction limit $V/g\rightarrow\infty$, as their intervening energy gaps diverge. 
 
%A finite difference $\Delta q_d$ between these eigenvalues corresponds to an 
%energy splitting $V\Delta q_d$ in the full Hamiltonian. 
%Consequently, in the limit $V/g\rightarrow\infty$, even a finite defect-induced splitting produces 
%a diverging energy gap, so that the resulting subspaces become
%gap protected.

Next we show how the single defect impacts the kinetic constraints, and hence the Hilbert-space connectivity. 
At the dynamical points considered in our work, the detuning can be decomposed as ($R$ is the maximum interaction range)
\begin{equation}
m=\sum_{r=1}^{R}m_r,
\qquad
m_r\in
\left\{
0,\frac{k_r}{2},k_r
\right\},
\end{equation}
with different values of $m_r$ corresponding respectively to blockade type, Ising type, and antiblockade type interactions
at range $r$. In general, the kinetic constraints associated with different interaction ranges must be satisfied simultaneously, such that an allowed local spin flip at site $i$ must preserve the defect-modified $q_{d,r}$ for every present interaction range $r$. For a given range $r$, the change in $q_{d,r}$ under a spin flip at site $i$ is
\begin{equation}
\Delta q_{d,r}
=
(1-2n_i)
\left[
\left(k_r+\delta_{i-r,i}\right)n_{i-r}
+
\left(k_r+\delta_{i,i+r}\right)n_{i+r}
-
2m_r
\right].
\label{eq:delta_qdr}
\end{equation}
%\revdel{The corresponding local spin-flip condition is therefore}
%\revdel{which is the defect-modified kinetic constraint. Similar to the cases without defects, the constraint depends on the interaction type (or value of $m_r$).}
If site $i$ is far from the defect, both  
$\delta_{i-r,i}$ and $\delta_{i,i+r}$ vanish, and the spin-flip condition $\Delta q_{d,r}=0$ becomes the same as that in the absence of the defect. 
If site $i$ is close to the defect, one of the two range-$r$ bonds connected to it is modified by the defect, which changes the spin-flip condition. 
In the following, we focus on this latter case and, without loss of 
generality, take $\delta_{i-r,i}=\delta_r$ and $\delta_{i,i+r}=0$. 
The spin-flip condition for bonds modified by the defect, or the kinetic constraint thereof, then becomes
\begin{equation}
\left(k_r+\delta_{r}\right)n_{i-r}
+k_r n_{i+r}
=2m_r.
\label{eq:defect_constraint}
\end{equation}

For the blockade type interaction with $m_r=0$, the kinetic constraint becomes
\begin{equation}
\left(k_r+\delta_r\right)n_{i-r}
+
k_rn_{i+r}
=0,
\end{equation}
which is only satisfied for $n_{i-r}=n_{i+r}=0$, the same as that in the absence of the defect. 
Thus, the Hilbert-space connectivity is not affected by the defect when $m=0$ (that is, 
interactions at all ranges are of the blockade type).

The situation is qualitatively different when $m\neq 0$, that is when $m_r\neq 0$ at some range $r$.
In this case, Eq.~(\ref{eq:defect_constraint}) is generally not satisfied, leading to suppressed local dynamics near the defect, where local spin flips are forbidden.
% \revadd{In this case, some spin flips allowed in the system without the defect no longer satisfy Eq.~(\ref{eq:defect_constraint}), thereby reducing the local connectivity near the defect.}
%Hence, for both Ising ($m_r=k_r/2$) and antiblockade ($m_r=k_r$) interactions, local defects generically interrupt the connectivity at the defect. 

% \revadd{For an Ising-type interaction with $m_r=k_r/2$, both $(n_{i-r},n_{i+r})=(1,0)$ and $(0,1)$ satisfy the spin-flip constraint without the defect. In the presence of the defect, Eq.~(\ref{eq:defect_constraint}) allows only $(n_{i-r},n_{i+r})=(0,1)$. Thus, the defect suppresses one of the two local spin-flip processes and reduces the connectivity near the defect.}
% \revadd{For an antiblockade-type interaction with $m_r=k_r$, the spin-flip constraint without the defect requires $(n_{i-r},n_{i+r})=(1,1)$. This configuration no longer satisfies Eq.~(\ref{eq:defect_constraint}). Away from the fine-tuned point $\delta_r=k_r$, no neighboring configuration satisfies the constraint, and local spin flips near the defect are forbidden.}

Nevertheless, we note that coincidental connectivity can emerge upon fine-tuning the defect. For example, when \(\delta_r=k_r\) for \(m_r=k_r\), Eq.~\eqref{eq:defect_constraint} is satisfied for \(n_{i-r}=1\) and \(n_{i+r}=0\).
%A similar scenario also arises at \(m_r=k_r/2\). 
These observations suggest that precise control of the atom geometry can be used to engineer more intricate patterns of Hilbert-space fragmentation.
% \revadd{Thus, the single geometric defect leaves blockade-type connectivity unchanged but modifies the Ising-type and antiblockade-type constraints.}

We illustrate the situations above in Fig.~\ref{fig:defect_q_subspace} with examples featuring a single defect.
We consider only nearest-neighbor interactions and introduce a single defect on the bond $(1,2)$ by setting $\delta_{1,2}=\delta$, while $\delta_{j,j+1}=0$ for all $j\neq1$. In Fig.~\ref{fig:defect_q_subspace}(a), at $m=0$, the $q=0$ subspace is robust: neither the interaction eigenvalue nor the 
connectivity changes.
In Fig.~\ref{fig:defect_q_subspace}(b), also at $m=0$, the $q=k_1$ subspace splits into $q_{\mathrm d}=k_1$ and $q_{\mathrm d}=k_1+\delta$ sectors, while its connectivity remains unchanged.
In Fig.~\ref{fig:defect_q_subspace}(c), at $m=k_1/2$, 
the original $q=k_1$ subspace (all product states are dynamically connected) also splits, while the local spin-flip rule is 
simultaneously modified, thereby reshaping the connectivity. 

The above discussion of a single defect can be extended to cases with multiple defects. At the dynamical point with $m=0$, the connectivity remains robust. At other dynamical points, each defect suppresses some local spin flips in its vicinity. Increasing the number of defects therefore further reduces the Hilbert-space connectivity. When the number of defects becomes sufficiently large to form spatial disorder, the situation changes qualitatively. Except for the robust \(m=0\) point, the spin-flip condition is broken at most locations, so that only local spin-flip processes survive. In this regime, the dynamical points no longer exhibit qualitatively different connectivity from generic detunings.

\section{experimental Implication}
Based on the analytical results above, a wide range of parameter and initial-state-dependent many-body dynamics are experimentally accessible by engineering the desired kinetic constraints.

More specifically, let us rewrite the Hamiltonian as
\begin{equation}
    \hat H=g\sum_i\hat X_i+\sum_{i,r}V_r\hat n_i\hat n_{i+r}-\Delta\sum_i\hat Z_i,
    \qquad V_r=k_rV.
\end{equation}  
In our work, the detuning is typically comparable to the interaction strength $V$, with $V\sim\Delta \gg g$. This parameter regime is experimentally accessible by choosing a sufficiently small $g$, whereas the interaction strength and range can be tuned by varying the geometry of the atom array or by dressing the Rydberg states.

In general, while dynamics are restricted to different $q$-subspaces, the presence of subspace fragmentation imposes further constraints on the dynamics. 
According to our analytical results, the subspace dynamics are frozen (that is, observables do not evolve in time) except for a discrete set of dynamical points, whereat the fragmentation structures of the relevant $q$-subspace are dictated by the emergent kinetic constraints, which in turn are dependent on the nature of the dynamical points. Here, by tuning the parameter $m=\Delta/V$ as well as the interaction range, one is able to control the nature of the relevant dynamical points (that is, the values of the constituting $m_r$), and the desired kinetic constraints can be realized according to the analytical insights provided by the auxiliary-fermion representation.
This opens up the avenue of engineering a wide range of constrained models in different $q$-subspaces, as exemplified by the PXP model (and its variants) and the integrable model Eq.~(9) as we have discussed in the main text. 
Furthermore, by initializing the system in different fragments, different dynamics, such as persistent oscillations, frozen dynamics, and subspace thermalization can all be realized, even as the Hamiltonian parameters are fixed. 

An outstanding prediction of our work is the Thomae-function characterization of the minimum spectral gap. Such an exotic phenomenon can in principle be detected using the many-body Ramsey interferometry developed specifically for Rydberg-atom arrays~\cite{Zeiher2016}, as outline below.

After initializing the system in the state $\ket{0}^{\otimes L}$, one applies a global $\pi/2$ pulse, to prepare the state
\begin{equation}
    \ket{+}^{\otimes L}
    =
    \frac{1}{\sqrt{2^L}}
    \sum_{\boldsymbol n}
    \ket{\boldsymbol n},
\end{equation}
where $\ket{+}=(\ket{0}+\ket{1})/\sqrt{2}$ and $\ket{\boldsymbol n}=\ket{n_1,n_2,\ldots,n_L}$ denotes the product-state basis. One then switches off the transverse Rabi coupling, $g=0$, so that the system evolves under
\begin{equation}
    \hat H_0=V\hat Q.
\end{equation}
Since $\hat Q$ is diagonal in the product-state basis, each $\ket{\boldsymbol n}$ acquires a phase determined by its eigenvalue $q_{\boldsymbol n}$. After an evolution time $t$, the state becomes (setting $\hbar=1$)
\begin{equation}
    \ket{\psi(t)}
    =
    \frac{1}{\sqrt{2^L}}
    \sum_{\boldsymbol n}
    e^{-iVq_{\boldsymbol n}t}
    \ket{\boldsymbol n}.
\end{equation}
A reverse global $\pi/2$ pulse is then applied, before the probability $P_{00\cdots0}(t)$ to find all atoms in the ground state at time $t$ is measured through the site-resolved fluorescence imaging. Since this probability is
\begin{align}
    P_{00\cdots0}(t)
    &=
    \left|
    \langle+|^{\otimes L}
    e^{-iV\hat Qt}
    |+\rangle^{\otimes L}
    \right|^2
    \nonumber\\
    &=
    \frac{1}{4^L}
    \sum_{\boldsymbol n,\boldsymbol n'}
    e^{-iV(q_{\boldsymbol n}-q_{\boldsymbol n'})t},
\end{align}
its Fourier spectrum contains all pairwise level spacings of $\hat Q$.  The Fourier peaks occur at
\begin{equation}
    \omega_{\boldsymbol n,\boldsymbol n'}
    =
    V\left|
        q_{\boldsymbol n}-q_{\boldsymbol n'}
    \right|,
\end{equation}
from which one can read off the minimum spectral gap
\begin{equation}
    \omega_{\min}
    =
    V\Delta q_{\min}.
\end{equation}
Repeating this measurement while varying the detuning should give the Thomae-function profile without the need to resolve each many-body energy level.

\end{document}